%% file: main.tex
\PassOptionsToPackage{table}{xcolor}
\documentclass[sigconf]{acmart}
\input{include/package}
\AtBeginDocument{%
  }

\setcopyright{none} 
\copyrightyear{2025}
\acmYear{2025}
\acmDOI{3695053.3731055}

\acmConference['ISCA 2025']{The 52nd IEEE/ACM International Symposium on Computer Architecture}{June 21--25, 2025}{Tokyo, Japan}
\acmISBN{978-1-4503-XXXX-X/18/06}

\begin{document}
\input{include/command}
\input{include/shorthands}

\title{FRED: A Wafer-scale Fabric for 3D Parallel DNN Training}
\input{include/author}
\ccsdesc[500]{Hardware~Emerging technologies}
\ccsdesc[500]{Networks~Network architectures}
\ccsdesc[500]{Computer systems organization~Architectures}

\renewcommand{\shortauthors}{Rashidi et al.}

\input{content/0_abstract}
\keywords{distributed training, wafer-scale platforms}

\maketitle
\input{content/1_introduction}
\input{content/2_background}
\input{content/3_motivation}
\input{content/4_design}
\input{content/5_routing}
\input{content/6_architecture}
\input{content/7_methodology}

\input{content/8_results}
\input{content/9_related}
\input{content/10_conclusion}

\begin{acks}
This work was supported by awards from Intel and the Semiconductor Research Corporation~(SRC).
Also, this research is supported by the ACE Center for Evolvable Computing, one of the seven SRC JUMP 2.0 centers.
We would like to thank the reviewers for their insightful comments. We also thank Jinsun Yoo for his help in revising the paper.  
\end{acks}

\bibliographystyle{ACM-Reference-Format}
\bibliography{reference/references}

\end{document}

%% file: include/package.tex
\usepackage{algorithm}
\usepackage{algpseudocode}
\usepackage{comment}
\usepackage{amsmath,amsfonts}
\usepackage{graphicx}
\usepackage{textcomp}
\usepackage{xspace}
\usepackage{fancyhdr}
\usepackage{subcaption}

\usepackage{enumitem}
\usepackage{soul}  

%% file: include/command.tex
\newcommand{\todo}[1]{\textcolor{red}{[TODO: #1]}}
\newcommand{\SR}[1]{\textcolor{blue}{[SR: #1]}}
\newcommand{\WW}[1]{\textcolor{violet}{[WW: #1]}}
\newcommand{\TK}[1]{\textcolor{red}{[TK: #1]}}
\newcommand{\PG}[1]{\textcolor{magenta}{[PG: #1]}}

\newcommand{\ParaTitle}[1]{\noindent\textbf{#1}}

\newcommand{\rom}[1]{\romannumeral #1}
\newcommand{\Rom}[1]{\textsc{\romannumeral #1}}

\newcommand{\RevTitle}[1]{\vspace{0.5em}\noindent\textbf{\emph{#1}}}
\newcommand{\ItemTitle}[1]{\noindent\textbf{#1}}
\newcommand{\KeyChange}[1]{\textbf{\textcolor{purple}{#1}}}

\definecolor{colorRevA}{rgb}{0.5, 1.0, 0.83}
\definecolor{colorRevB}{rgb}{0.8, 1.0, 0.0}
\definecolor{colorRevC}{rgb}{0.88, 0.69, 1.0}
\definecolor{colorRevD}{rgb}{1.0, 0.6, 0.4}
\definecolor{colorRevE}{rgb}{0.7, 0.75, 0.71}
\colorlet{colorRevCom}{yellow}
\DeclareRobustCommand{\RevA}[1]{{\sethlcolor{colorRevA}\hl{#1}}}
\DeclareRobustCommand{\RevB}[1]{{\sethlcolor{colorRevB}\hl{#1}}}
\DeclareRobustCommand{\RevC}[1]{{\sethlcolor{colorRevC}\hl{#1}}}
\DeclareRobustCommand{\RevD}[1]{{\sethlcolor{colorRevD}\hl{#1}}}
\DeclareRobustCommand{\RevE}[1]{{\sethlcolor{colorRevE}\hl{#1}}}
\DeclareRobustCommand{\RevCom}[1]{{\sethlcolor{yellow}\hl{#1}}}

\newcommand{\insertFigureRev}[6]{
    \begin{figure}[t]
      \centering
      \includegraphics[width=#3\linewidth, cfbox=#6 2pt 2pt]{figs/#1}
      \vspace{#4}
      \caption{\small #2}
      \label{fig:#1}
      \vspace{#5}
    \end{figure}
}

\soulregister\cite7
\soulregister\autoref7
\soulregister\ref7
\soulregister\pageref7
\soulregister\ours7

\def\sectionautorefname{Section}
\def\subsectionautorefname{Section}
\def\subsubsectionautorefname{Section}
\def\figureautorefname{Figure}

\newcommand{\insertFigure}[5]{
    \begin{figure}[h]
      \centering
      \includegraphics[width=#3\linewidth]{figs/#1.pdf}
      \vspace{#4}
      \caption{\small #2}
      \vspace{#5}
      \label{fig:#1}
    \end{figure}
}

\newcommand{\insertFigureWide}[5]{
    \begin{figure*}[h]
      \centering
      \includegraphics[width=#3\linewidth]{figs/#1.pdf}
      \vspace{#4}
      \caption{\small #2}
      \vspace{#5}
      \label{fig:#1}
    \end{figure*}
}

%% file: include/shorthands.tex
\newcommand{\tool}{{\sc ASTRA-sim}~}

\newcommand{\resnet}{ResNet-152\xspace}
\newcommand{\transsmall}{Transformer-17B\xspace}
\newcommand{\gpt}{GPT-3\xspace}
\newcommand{\translarge}{Transformer-1T\xspace}

\newcommand{\ours}{\textsc{Fred}\xspace}
\newcommand{\oursns}{\textsc{Fred}}

%% file: include/author.tex
\author{Saeed Rashidi}
\authornote{The author is now at Meta: rashidi1saeed@meta.com}
\email{saeed.rashidi@gatech.edu}
\orcid{0000-0002-6472-9920}
\affiliation{%
  \institution{Georgia Institute of Technology}
  \streetaddress{North Ave NW}
  \city{Atlanta}
  \state{Georgia}
  \country{USA}
}

\author{William Won}
\email{william.won@gatech.edu}
\orcid{0000-0002-1715-9144}
\affiliation{%
  \institution{Georgia Institute of Technology}
  \streetaddress{North Ave NW}
  \city{Atlanta}
  \state{Georgia}
  \country{USA}
}

\author{Sudarshan Srinivasan}
\email{sudarshan.srinivasan@intel.com}
\orcid{0009-0002-8662-5820}
\affiliation{%
  \institution{Intel}
  \city{Bangalore}
  \state{Karnataka}
  \country{India}
}

\author{Puneet Gupta}
\email{puneetg@ucla.edu}
\orcid{0000-0002-6188-1134}
\affiliation{%
  \institution{UCLA}
  \city{Los Angeles}
  \state{California}
  \country{USA}
}

\author{Tushar Krishna}
\email{tushar@ece.gatech.edu}
\orcid{0000-0001-5738-6942}
\affiliation{%
  \institution{Georgia Institute of Technology}
  \streetaddress{North Ave NW}
  \city{Atlanta}
  \state{Georgia}
  \country{USA}
}

%% file: content/0_abstract.tex
\begin{abstract}

Wafer-scale systems are an emerging technology that tightly integrates high-end accelerator chiplets with high-speed wafer-scale interconnects, enabling low-latency and high-bandwidth connectivity.
This makes them a promising platform for deep neural network~(DNN) training.
However, current network-on-wafer topologies, such as 2D Meshes, lack the flexibility needed to support various parallelization strategies effectively.
In this paper, we propose \ours, a wafer-scale fabric architecture tailored to the unique communication needs of DNN training.
\ours creates a distributed on-wafer topology with tiny microswitches, providing nonblocking connectivity for collective communications between arbitrary groups of accelerators and enabling in-switch collective support.
Our results show that for sample parallelization strategies, \ours can improve the average end-to-end training time of \resnet, \transsmall, \gpt, and \translarge by 1.76$\times$, 1.87$\times$, 1.34$\times$, and 1.4$\times$, respectively, compared to a baseline wafer-scale Mesh.

\end{abstract}

%% file: content/1_introduction.tex
\section{Introduction}

DNN models are on an exponential growth curve.
A recent study shows that in less than two years, the compute and memory requirements for DNN training have increased by 1,800$\times$ and 1,500$\times$, respectively~\cite{CerebrasKeynote}.
Distributing the training across multiple accelerators or neural processing units~(NPUs) is a common practice today to reduce the training time.
However, one critical side effect of distributed training is the communication overhead between NPUs to synchronize model gradients and/or activations, depending on the parallelization strategy.
As the number of NPUs scales, communication overhead increases, up to a point where it becomes the dominant factor in distributed training latency~\cite{astrasim,hotiPaper,FlexFlow,NVidiaSwitch}.

There are fundamental limits to the bandwidth that can be provided even by high-speed rack-scale fabrics (such as NVLink~\cite{NVlink}), and thus there has been a growing interest in platforms that integrate multiple NPUs together in the same package.
Cerebras~\cite{Cerebras} demonstrated one extreme incarnation of this idea in the form of a monolithic wafer with NPUs connected to one another.
More cost- and yield-effective approaches include silicon/organic interposer-based approaches~\cite{Simba,hwang2020centaur} or using Silicon Interconnect Fabric~(Si-IF), which bonds chiplets directly onto a full thickness silicon \emph{wafer} without needing a package~\cite{waferScaleGPU, WaferScaleChiplets,WaferScaleSwitches}.
\emph{In this work, we assume a passive, interconnect-only wafer-scale substrate onto which chiplets are bonded at fine pitch similar to Si-IF or TSMC-SoW~\cite{tsmc-sow}.
This allows for heterogeneous integration of compute, memory, and network chiplets from disparate technologies, unlike the monolithic Cerebras approach.}

While there is broad agreement on the scalability and bandwidth benefits that wafer-scale substrates can provide, the \emph{architecture of the fabric connecting the NPUs} remains an open question.
All wafer-scale accelerator proposals to date (e.g., Cerebras CS2~\cite{Cerebras}, NVIDIA's SIMBA~\cite{Simba}, UCLA's waferscale GPU~\cite{waferScaleGPU}, Chiplet Cloud~\cite{chipletcloud}, Chen et al., TTO~\cite{tto}) have implemented a 2D Mesh topology for the fabric.
The choice of a Mesh is understandable.
It is the most pervasive topology in many-core chips given its ease of place-and-route and scalability and is the natural choice on a wafer-scale substrate as well.
\emph{However, we demonstrate that the inherent blocking nature of the 2D Mesh topology is extremely inefficient for DNN training communication use cases}.


\begin{figure}[h]
  \centering
  \includegraphics[width=1\linewidth]{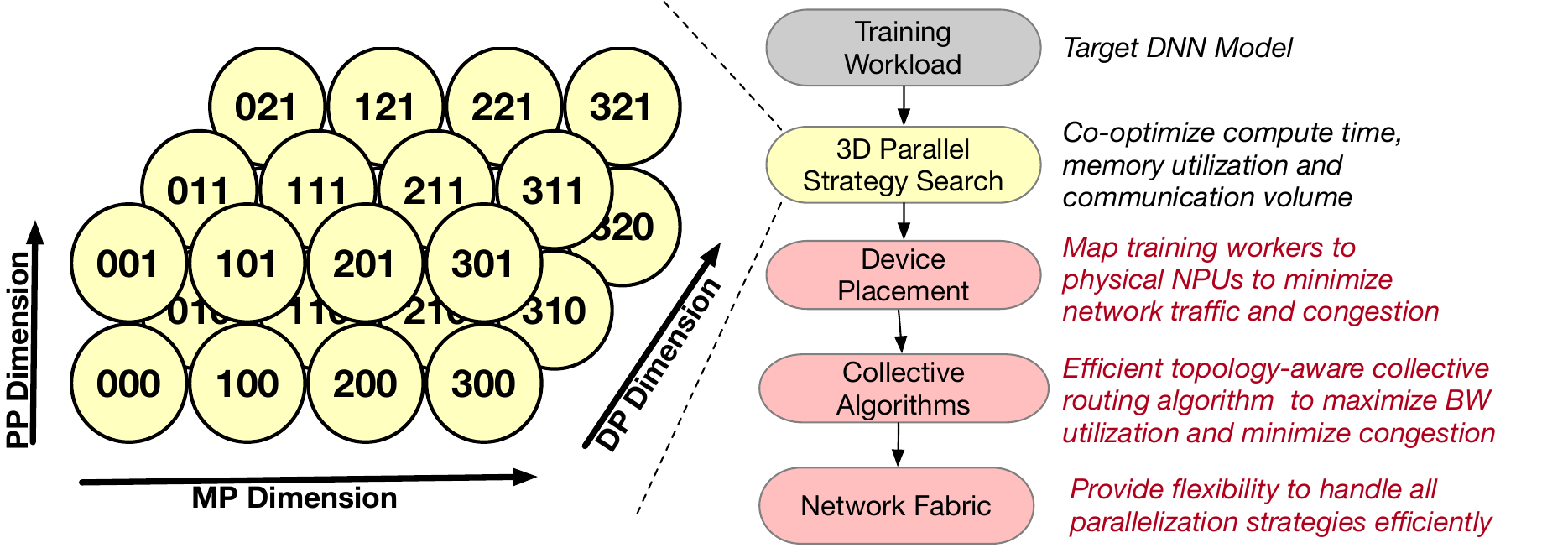}
  \vspace{-2em}
  \caption{\small HW-SW Co-Design Stack for Optimizing DNN training communication.
    This work addresses the three phases highlighted in red.
    The left part of the figure shows a sample logical view of training workers in 3D parallelism.
    The parallelization strategy is of size MP(4)-DP(3)-PP(2), meaning that there are 4/3/2 peer workers for the MP/DP/PP dimension.
    Each worker is named with 3 digits, representing the ID of the worker in the MP, DP, and PP dimensions, respectively.
    Workers that are aligned along each dimension should \emph{communicate} for that respective dimension's parallelization type.
    For example, workers 000, 100, 200, and 300 should communicate for MP type (i.e., activation/input gradient sync during forward-pass/back-propagation), while workers 300, 310, 320 should communicate for DP type (weight gradient sync during backpropagation).}
    \vspace{-1.5em}
  \label{fig:3dparallel_fredoverview}
\end{figure}

Communication in DNN training depends inherently on the parallelism strategy being employed.
Data-Parallel (DP)~\cite{PytorchDistributedDP,NVidiaSwitch}, Model-Parallel (MP)~\cite{GShard,FlexFlow}, and Pipeline-Parallel (PP)~\cite{GPipe,Pipedream} are the building blocks of any parallelism strategy.
In DP, the DNN model is replicated across NPUs and each NPU works on a different set of training samples~(i.e., minibatch).
In MP, each DNN layer is sharded across NPUs while they work on the same training samples.
In PP, each NPU hosts a subset of DNN layers, and training samples flow through the NPUs in a pipeline manner.
3D parallelism~\cite{3DParallelMegatron} utilizes all the aforementioned strategies by creating different MP/DP/PP groups between NPUs.
The optimal balance between DP, MP, and PP is heavily dependent on the workload and underlying platform and can \emph{significantly vary for different workload/platform configurations}~\cite{FlexFlow,alibaba_hpn}.
\autoref{fig:3dparallel_fredoverview} shows an example of a 3D parallelism strategy.


\begin{figure}[h]
  \centering
  \includegraphics[width=0.85\linewidth]{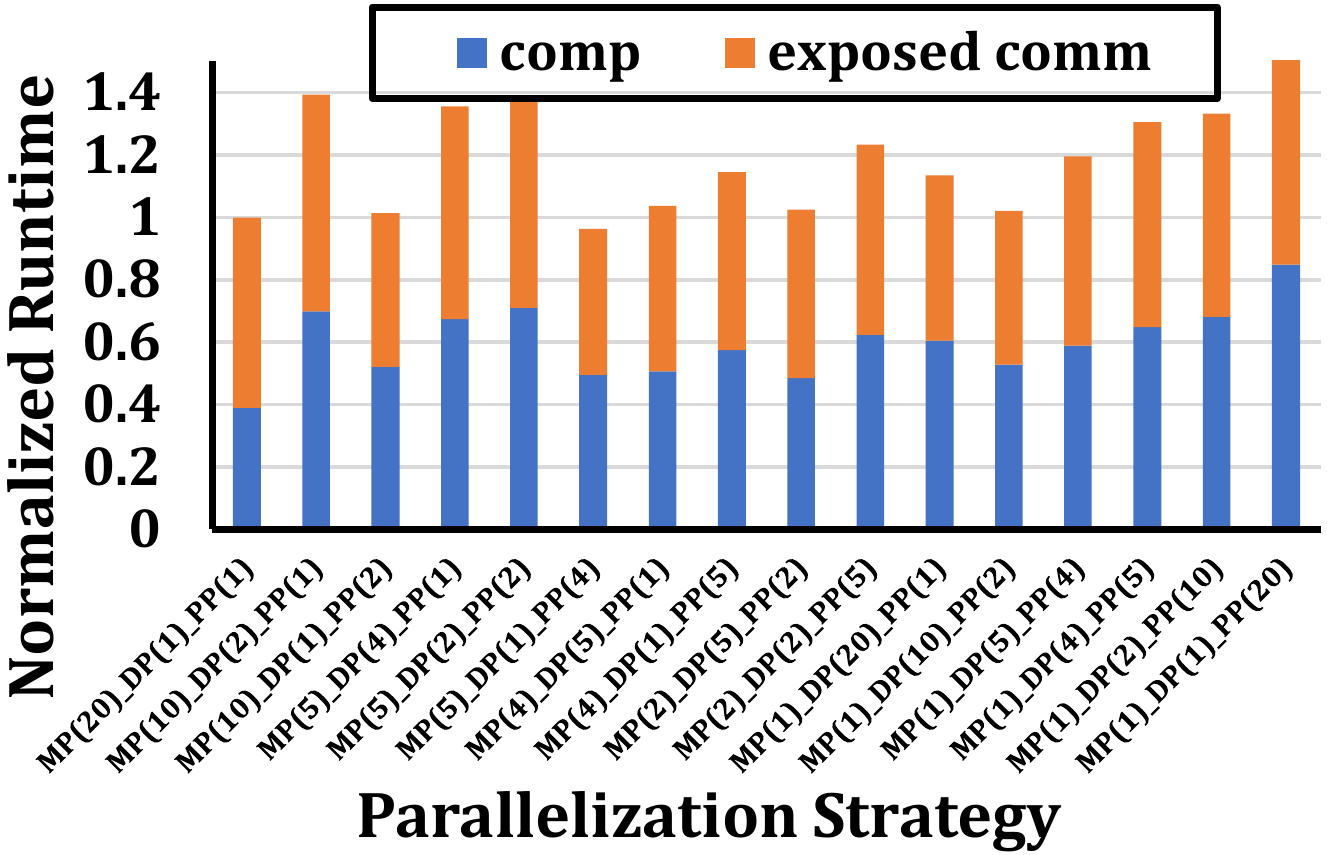}
  \vspace{-1em}
  \caption{\small The normalized computation/communication overhead for various parallelization strategies (explained in~\autoref{sec:background}) of \transsmall, when running on top of the 2D Mesh topology (as described in~\autoref{sec:methodology}) that connects 20 NPUs on the wafer.}
  \vspace{-1.5em}
  \label{fig:Motiv}
\end{figure}

From a communication perspective, 3D parallelism requires executing \emph{multiple concurrent communication operations} between NPUs within the same MP/DP/PP group at different stages of distributed training.
Moreover, different parallelism strategies stress the compute and communication differently.
This is quantified in~\autoref{fig:Motiv}.
As the figure shows, high communication overhead can result in the total training overhead of compute-efficient strategies being greater than that of less compute-efficient strategies (e.g., MP(20)-DP(1)-PP(1) vs. MP(5)-DP(4)-PP(1)).
Other than the communication volume which is determined by the workload, the main purpose of such high network overhead in~\autoref{fig:Motiv} is the \emph{inefficient use of network resources} in the baseline topology. This is mainly because: (i) for the majority of the comm operations, only half or less than half of the NPU links get activated (discussed in detail in \autoref{subsub:BWUtilizationChallenge}), (ii) network contention between MP/DP/PP parallel groups (discussed in detail in \autoref{subsub:devicePlacementChallenge}). The full list of baseline topology challenges is discussed in \autoref{sec:challenges_mesh}.

In summary, an optimal wafer-scale fabric for distributed DNN training should meet the following three needs:
 \begin{enumerate}
     \item Handle \emph{multiple} non-blocking collective communications with minimum congestion.
     \item Be efficient for \emph{all} 3D parallelism configurations.
     \item Provide \emph{high-BW} connectivity between NPUs.
 \end{enumerate} 

In this work, we propose \textbf{\ours}, a wafer-scale fabric with Flexible REduction-Distribution feature for supporting arbitrary 3D parallelism.
\ours includes: (\rom{1})~a novel topology with switches that provide native support for reduction and broadcast for bandwidth amplification, (\rom{2})~a collective routing algorithm with non-blocking support, and (\rom{3})~a device placement algorithm to minimize congestion.
We deploy \ours over a wafer-scale substrate~\cite{WaferScaleChiplets}.
Each NPU in our architecture is a hybrid integration of high-end compute chiplets and 3D-stack DRAM chiplets (analogous to H100~\cite{NvidiaH100}).
We also discuss solutions to physically layout and scale the \ours topology over a wafer substrate.

To the best of our knowledge, \emph{\ours is the first wafer-scale fabric proposal tailored for DNN training, that can efficiently support multiple concurrent collectives for hybrid parallelization strategies (e.g., 3D parallelism)}.
Hence, \ours enables the compiler to consider any type of parallelization strategy without any concern about how efficiently they can be executed on the network. To summarize:
\begin{itemize}
    \item We motivate the challenges with designing a wafer-scale fabric for 3D parallelism~(\autoref{sec:motivation}).
    \item We propose \ours, a novel network fabric that includes several innovative features: \emph{a switch fabric} with flexible reduction-distribution trees connected via a scalable topology (\autoref{sec:design}), and \emph{a novel routing algorithm} to route multiple collectives concurrently, along with a congestion-aware device placement policy for 3D parallelism~(\autoref{sec:fred_routing}).
    \item We demonstrate how \ours can be implemented as a wafer-scale fabric~(\autoref{sec:methodology}).
    \item We compare \ours with baseline fabrics for some sample workloads and parallelization strategies~(\autoref{sec:results}).
\end{itemize}

Our results show that \ours can improve the average end-to-end training time of \resnet, \transsmall, \gpt, and \translarge by 1.76$\times$, 1.87$\times$, 1.34$\times$, and 1.4$\times$, respectively, when compared to the baseline 2D Mesh.

%% file: content/2_background.tex
\vspace{-2mm}
\section{Background}\label{sec:background}

\subsection{Collective Communication Patterns}
\label{subsec:collectivepatterns}

\begin{figure}[h]
  \centering
  \includegraphics[width=0.95\linewidth]{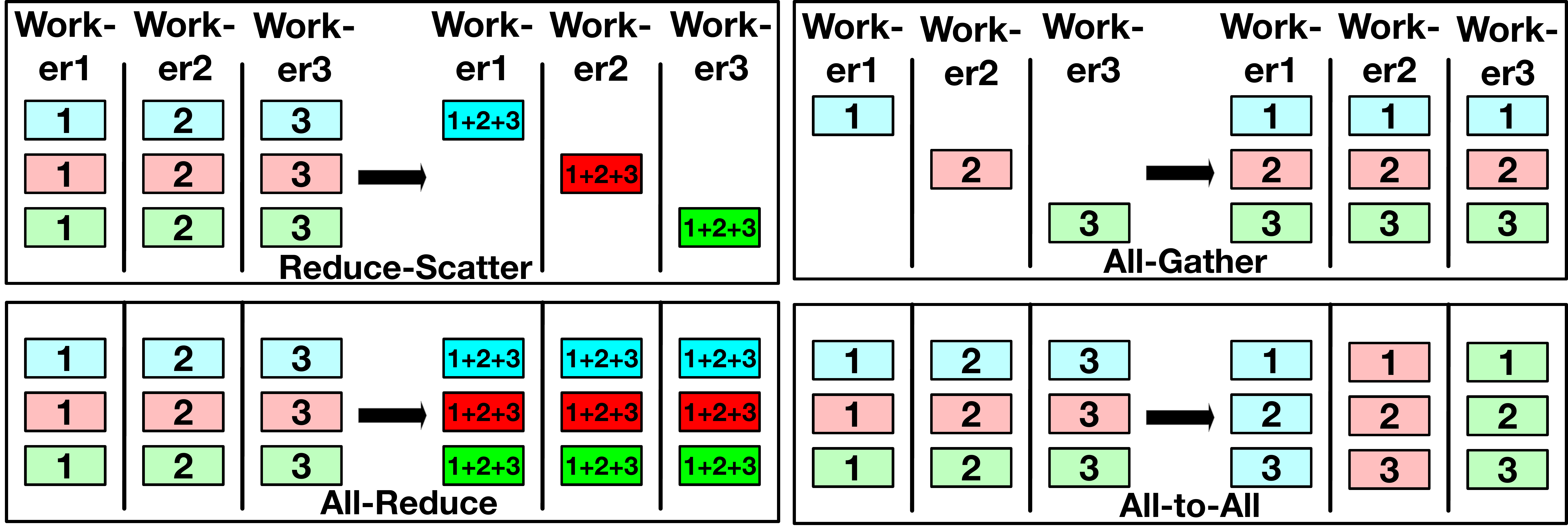}
  \vspace{-1em}
  \caption{\small Collective communication patterns among three workers.}
  \vspace{-1em}
  \label{fig:Collective}
\end{figure}

Although DNN models can be highly diverse, most of their communication during distributed training can be handled through collective patterns~\cite{NVidiaSwitch}.  
Depending on the model type and parallelization strategy, different types of collectives may be needed to synchronize on activations/gradients during forward-pass/back-propagation~\cite{astrasim}.  
\autoref{fig:Collective} shows the mathematical implication of the most common collective patterns between three workers.  
During \emph{Reduce-Scatter}, workers communicate in such a way that, at the end, each worker has a portion of globally reduced data.  
In \emph{All-Gather}, each worker broadcasts its local data to all other workers.  
\emph{All-Reduce} is the most common pattern in distributed training~\cite{NVidiaSwitch} and can be thought of as a \emph{Reduce-Scatter} followed by an \emph{All-Gather}.  
In \emph{Reduce}, multiple NPUs participate in reducing data, and the result is stored only on one NPU, while \emph{Gather} collects the data from all NPUs and stores them on a single NPU. \emph{Multicast} means a single NPU sends its data to multiple NPUs.  
In \emph{All-to-All}, each worker sends a portion of its local data to each worker.

\vspace{-3mm}

\subsection{Collective Communication Algorithms}\label{subsec:collectivealgs}

The patterns described in~\autoref{subsec:collectivepatterns} can be handled through different \emph{collective algorithms}. In general, there are two distinct way to implement such algorithms:

\ParaTitle{1) Endpoint-based.}
NPUs communicate in a peer-to-peer distributed manner through explicit send/recv of messages with themselves and without requiring central coordination. In this case,  
the optimal algorithm is usually dependent on the physical network topology and collective size.
For example, ring-based All-Reduce is optimal when the physical topology is a ring, while tree-based All-Reduce is optimal for tree-based topologies, or when the message size is small~\cite{collective1}. 

One drawback of the NPU-to-NPU-based approach is the amount of traffic it generates.  
For example, the most BW-optimal NPU-to-NPU algorithms require each NPU to send/receive nearly $\frac{2(N-1)}{N}D$ bytes of data to execute an All-Reduce of $D$ bytes among $N$ NPUs~\cite{NVidiaSwitch}, which is almost $2\times$ of the All-Reduce size ($D$ bytes)~\cite{collective1,collective2,collective10}.
This is because all endpoint-based algorithms must perform reduction and gather phases separately, resulting in $\frac{(N-1)}{N}D$ send/recv per NPU to accomplish each phase, respectively~\cite{collective1,collective2,collective10}.  

\ParaTitle{2) In-Network Collective Execution.}
To alleviate the extra traffic of endpoint-based approach, recent proposals have introduced in-network collective algorithms by adding compute capability to the switches~\cite{switchml,NVidiaSwitch,CommBottleneck1} to perform both reduction and gather at the same time.   
For example, an All-Reduce of $D$ bytes only requires each NPU to send/receive $D$ bytes to the switch/switch-hierarchy. The switch/switch-hierarchy receives $D$ bytes from each NPU, performs reduction across all received data from all $N$ NPUs, and broadcasts $D$ bytes back to all NPUs. Therefore, compared to the endpoint-based approach, each NPU sends/receives almost half the traffic ($D$ bytes vs. $\frac{2(N-1)}{N}D$ bytes.)~\cite{CommBottleneck1}. Additionally, In-Network collective execution allows the endpoint resources to be allocated for training compute tasks, while the network switches handle the collectives efficiently. 
\subsection{Communication in 3D Parallelism}\label{subsec:3DPara}

There are multiple ways of distributing the distributed training tasks across multiple NPUs (a.k.a. the parallelization strategy): MP (a.k.a. Tensor-parallelism)~\cite{megatronlm}, DP~\cite{PytorchDistributedDP}, and PP~\cite{GPipe,Pipedream}.  
The combination of these strategies can be generalized in the form of 3D-parallelism~\cite{3DParallelMegatron}.  
\autoref{fig:3dparallel_fredoverview} shows the concept of 3D-parallelism.  
In this case, each training worker is part of one MP, DP, and PP group, where the ID (offset) of each NPU within its MP/DP/PP group is determined using the first/second/third digits of a 3-digit worker ID.  
Therefore, the NPUs that have the same DP \& PP digits are within the same MP group (e.g., 000, 100, 200, and 300).  

The NPUs within the same DP group should communicate through the \emph{All-Reduce} collective pattern during back-propagation to sync their locally computed model gradients and update the model before starting the next training iteration~\cite{NVidiaSwitch}.  
For the MP group case, NPUs need to communicate during forward-pass/back-propagation to synchronize on output-activations/input-gradients.  
The communication pattern, however, depends on the layer type and the way it is sharded.  
The usually observed patterns are: \emph{All-Reduce}~\cite{megatronlm}, \emph{All-to-All}~\cite{DLRM}, \emph{Reduce-Scatter}~\cite{GShard}, and \emph{All-Gather}~\cite{GShard}.
For the PP group, the NPUs need to transfer the output-activations/input-gradients during forward-pass/back-propagation on the borderline layers and pass the data to the NPU(s) hosting the next set of layers. \autoref{table:coll_types} represents collective patterns incurred by each parallelization strategy.

\autoref{fig:3dparallel_fredoverview} also shows the necessity to handle multiple collectives at the same time.  
For e.g., there are eight different DP groups, meaning that up to eight concurrent All-Reduces should be handled for the DP communications (similarly, there are six/twelve concurrent communication operations for MP/PP communications).  
Moreover, the communication type and peer workers differ across MP, DP, and PP groups.  
Thus, \emph{it is crucial for the underlying network fabric to be flexible for concurrent and different collective patterns}.

\vspace{-4mm}
\input{tables/coll_types.tex}

\subsection{Multi-chiplet Integration}\label{subsec:waferscale}

In chiplet-based integration, NPU chips are fabricated and then bonded to a package interconnect (e.g., Si-IF)~\cite{Simba,WaferScaleChiplets,tsmc-sow}.  
In this approach, components from different technologies (e.g., even DRAM) can be integrated on the package.  
Additionally, since the chiplets can be tested before integration, this approach has a better yield, supports heterogeneity, and requires less redundancy compared to fully monolithic approaches such as Cerebras~\cite{Cerebras}.

\ParaTitle{Multi-chiplet Fabric Topologies.}  
Recent products and research in multi-chiplet platforms are all based on interconnecting NPUs through a 2D-mesh topology~\cite{waferScaleGPU,Cerebras,WaferScaleChiplets,Simba,tto,chipletcloud}.  
Among the main reasons for choosing 2D-mesh for on-package/on-wafer is ease of place \& route and area optimality over a 2D substrate~\cite{waferScaleGPU}.  
\textbf{\emph{Thus, in this paper, we choose 2D-mesh as the main baseline topology and compare our proposal against it}}.

%% file: tables/coll_types.tex
\begin{table}[h]
\small
\caption{\small Collective patterns incurred by distinct parallelizations.}
\label{table:coll_types}
\vspace{-1em}

\resizebox{0.85\columnwidth}{!}{

\begin{tabular}{cccccc}
\toprule
\textbf{Parallelism} & \textbf{\begin{tabular}[c]{@{}c@{}}Reduce-\\ Scatter\end{tabular}} & \textbf{\begin{tabular}[c]{@{}c@{}}All-\\ Gather\end{tabular}} & \textbf{\begin{tabular}[c]{@{}c@{}}All-\\ Reduce\end{tabular}} & \textbf{All-to-All} & \textbf{\begin{tabular}[c]{@{}c@{}}Point-to-\\ Point\end{tabular}} \\
\midrule
Model & \checkmark & \checkmark & \checkmark & \checkmark &  \\
Data &  &  & \checkmark &  &  \\
Pipeline &  &  &  &  & \checkmark \\
3D & \checkmark & \checkmark & \checkmark & \checkmark & \checkmark \\
\bottomrule
\end{tabular}

}

\vspace{-2em}
\end{table}

%% file: content/3_motivation.tex
\section{Desired Metrics for a Wafer-scale Fabric}\label{sec:motivation}


\subsection{Communication Demands}
\label{sec:execution_modes}

First, we discuss two execution modes for running DNN training over a wafer-scale substrate.

\subsubsection{Weight Stationary}
When DNN models can fit entirely in the available on-chip memory within a wafer, loading the entire model parameters and collecting the training result to/from the package is a one-time overhead.\footnote{
All model updates over different training iterations happen on-chip.}
The cost of loading the pre-trained and storing the trained model is amortized over thousands of training iterations. The input samples, however, need to be loaded at the beginning of each training iteration. Such I/O operations have minimal impact on the overall training performance since the samples are much smaller than the model size. Therefore, \emph{in this mode, the main performance factor is the efficiency of compute cores and the NPU-to-NPU communication performance}. A non-optimized interconnect can result in poor NPU-to-NPU communication performance for certain parallelization strategies, forcing the compiler to discard some strategies despite their better compute and on-chip memory utilization, solely because of their poor communication performance (\autoref{sec:challenges_mesh}).

\subsubsection{Weight Streaming} When the available on-chip memory is insufficient to fit the model, the execution model shifts to \emph{weight streaming}~\cite{Cerebras,CerebrasKeynote}. In this scenario, only a subset of DNN layers is loaded onto the package at any given time. After processing these layers, the on-chip storage is reclaimed for the next set of layers. Consequently, the entire model must be loaded onto the chip multiple times during the model training (at least once during the forward pass and once during back-propagation). Additionally, as NPUs compute model gradients, they push this data to off-chip storage, and a lightweight on-storage compute core updates the model for the next iteration\footnote{
Model updates involve low operational intensity. Hence, performing these updates off-chip prevents wasting I/O bandwidth by avoiding loading optimizer states onto the chip for lightweight operations.}~\cite{Cerebras}. This approach makes the performance I/O bound, meaning that the upper-bound training performance scales as $\propto \frac{model\_size}{I/O\_BW}$. Therefore, \emph{in addition to compute efficiency and NPU-to-NPU communication performance, maintaining maximum I/O bandwidth is crucial}. A rigid topology can create hotspots when distributing/collecting the model/gradients to/from the I/O channels, which \emph{limits the I/O data rate} (\autoref{sec:challenges_mesh}) and directly impacts training performance.

\subsection{Challenges with 2D Mesh}
\label{sec:challenges_mesh}

Next, we discuss specific challenges in a 2D Mesh for supporting the communication needs of DNN training.

\subsubsection{Efficient I/O}
\begin{figure}[h]
  \centering
  \includegraphics[width=0.8\linewidth]{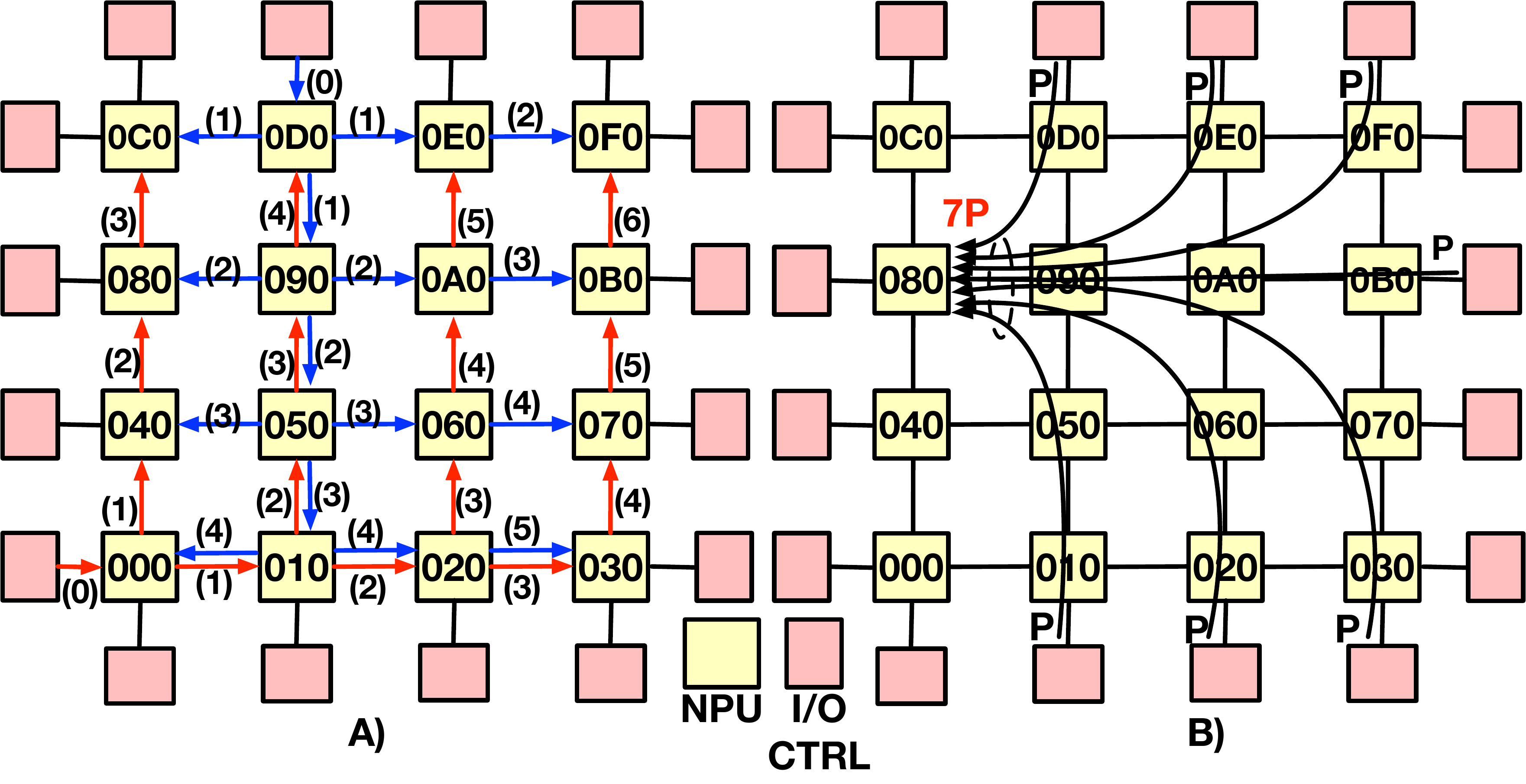}
  \vspace{-1em}
  \caption{\small (A) The broadcast communication pattern when reading from two different I/O channels (shown in red and blue arrows). The number associated with each arrow shows the timestamp when data crosses that link for one packet. In practice, multiple packets are pipelined across each path. In this example, the parallelization strategy is MP(1)-DP(16)-PP(1), and the model weights are broadcast among all NPUs for the weight streaming execution model. Note that the reverse order is used to sum the weight gradients during the back-propagation and write the final results into the remote storage. (B) The maximum channel load analysis corresponding to~\autoref{fig:DataIngestion}.a, when all of the I/O channels are used simultaneously.}
  \vspace{-1.5em}
  \label{fig:DataIngestion}
\end{figure}

As mentioned earlier, maintaining high I/O bandwidth is critical for achieving optimal performance in the weight streaming execution model. However, the 2D mesh often falls short of delivering maximum I/O performance. \autoref{fig:DataIngestion}~illustrates this problem using a $4\times 4$ mesh topology with a pure DP parallelization strategy. In this scenario, each weight fetched from an off-chip memory channel must be broadcast to all NPUs. \autoref{fig:DataIngestion}(A) shows a broadcast algorithm, based on the \emph{MPI} implementation of one-to-many pattern on 2D mesh~\cite{MPI}, when reading from two different memory channels (shown as the red and blue flows).

Ideally, all memory channels should stream (different) weights simultaneously and with the line-rate to maximize the I/O BW. However, \emph{the shape of 2D Mesh topology inherently generates hotspots and negatively affects the I/O BW}. \autoref{fig:DataIngestion}(B) shows the maximum channel load, for one hotspot link, when all memory channels are fetching the weights at the same time. If the BW of each memory channel is $P$ bytes/s, then the hotspot link should have the capacity (BW) of $7P$ bytes/s to allow the maximum I/O BW on a $4\times 4$ mesh.


In general, for an $N \times N$ mesh and $4 \times N$ external I/O channels, the wafer-scale fabric links should have a bandwidth of $\mathbf{(2N-1)P}$ bytes/s to fully utilize the I/O bandwidth in all parallelization strategies, assuming each I/O channel has a bandwidth of $P$ bytes/s. As the formula indicates, the required link bandwidth grows $O(N)$ with the mesh width. For larger packages, the technology might not support such high-bandwidth requirements on the package. In such cases, the I/O channel rate must be scaled down proportionally to accommodate the maximum link bandwidth, i.e., $P=\frac{link\_BW}{(2N-1)}$.

\ParaTitle{\ours's Solution.} \ours prevents network hotspots by adaptively routing the traffic through all of its links equally, enabling further scalability of the wafer-scale systems.

\subsubsection{Ease of Device Placement}\label{subsub:devicePlacementChallenge}


\begin{figure}[h]
  \centering
  \includegraphics[width=0.8\linewidth]{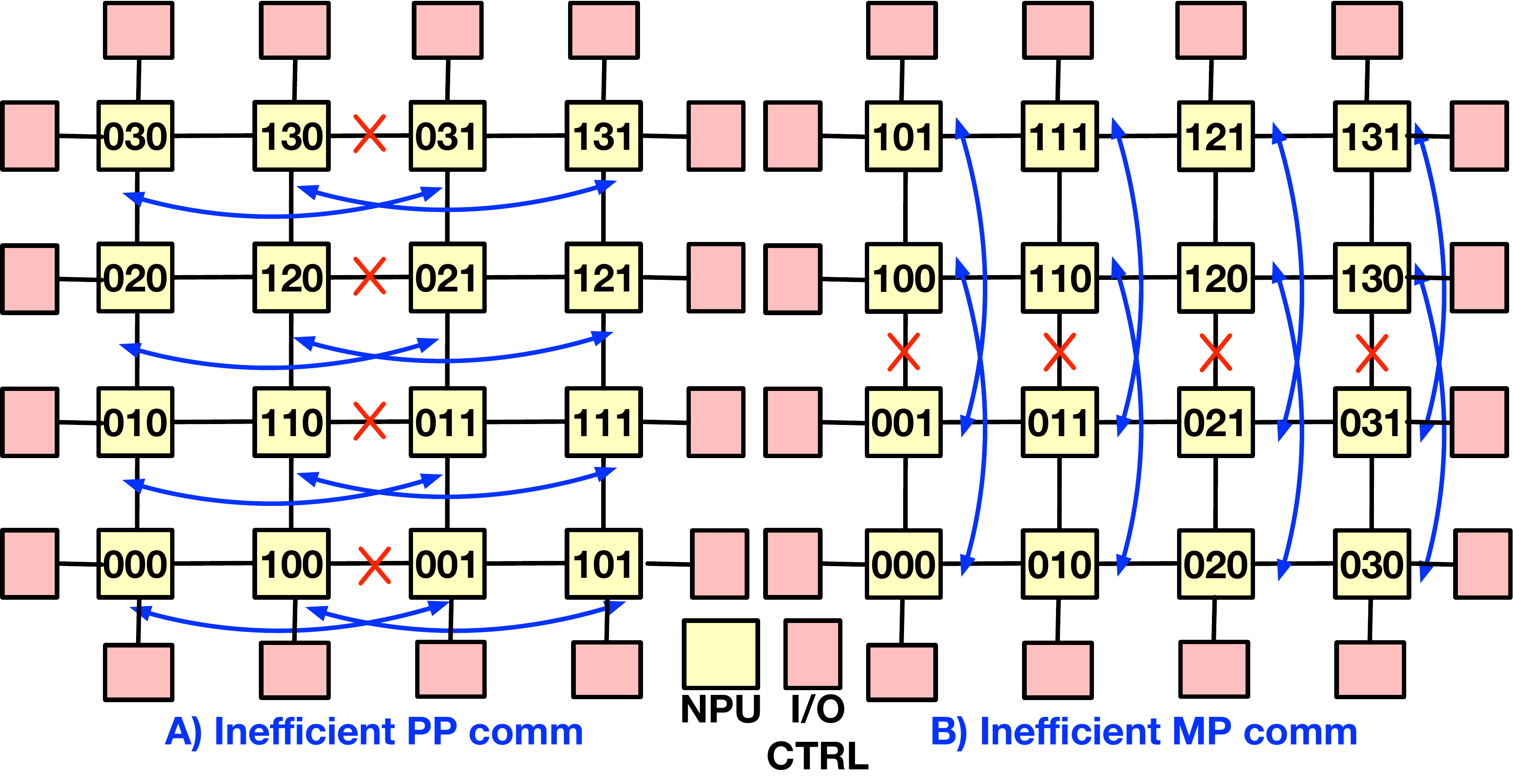}
  \vspace{-1em}
  \caption{\small Two different device placement mappings for an MP(2)-DP(4)-PP(2) strategy. (A) A device placement that favors MP and DP communications but causes congestion for PP communications. (B) A device placement that favors DP and PP communications but causes congestion for MP communications.}
  \vspace{-1.5em}
  \label{fig:DevicePlacement}
\end{figure}

Device placement involves assigning each logical training worker to a physical NPU. With $N$ NPUs, there are $N!$ possible device placement mappings. This becomes critical in 3D parallelism, as each training worker may have different communication volumes and patterns with other workers across distinct parallelization groups (refer to~\autoref{fig:3dparallel_fredoverview}). Therefore, finding a device placement that minimizes network contention is essential.

However, this is challenging with rigid topologies, especially 2D Mesh, where certain communication patterns are inherently prioritized over others. \autoref{fig:DevicePlacement}~illustrates two different mappings for a given MP(2)-DP(4)-PP(2) strategy. In~\autoref{fig:DevicePlacement}(A), the MP and DP communications are free of congestion, but PP communications cause congestion between different PP groups. Conversely, in~\autoref{fig:DevicePlacement}(B), DP and PP communications are optimized, but MP communications face congestion between MP groups.
Ultimately, as 2D mesh offers two logically disjoint dimensions ($x$ and $y$), \emph{it is mathematically impossible for all 3D parallelism dimensions to be optimally mapped onto a 2D Mesh}. This is trivial by observing the four corner NPUs, where each NPU offers two outgoing links. Consequently, due to the limited path diversity, one out of the three parallelization groups must experience network congestion and reduced communication performance. 
Determining which communication patterns to prioritize, unavoidable on 2D Mesh, requires a thorough analysis of the end-to-end workload and understanding the impact of different communication operations.

\ParaTitle{\ours's Solution.} \ours supports congestion-free routing for all communication patterns simultaneously.

\subsubsection{Non-Aligned Parallelization Strategies}


\begin{figure}[h]
  \centering
  \includegraphics[width=0.8\linewidth]{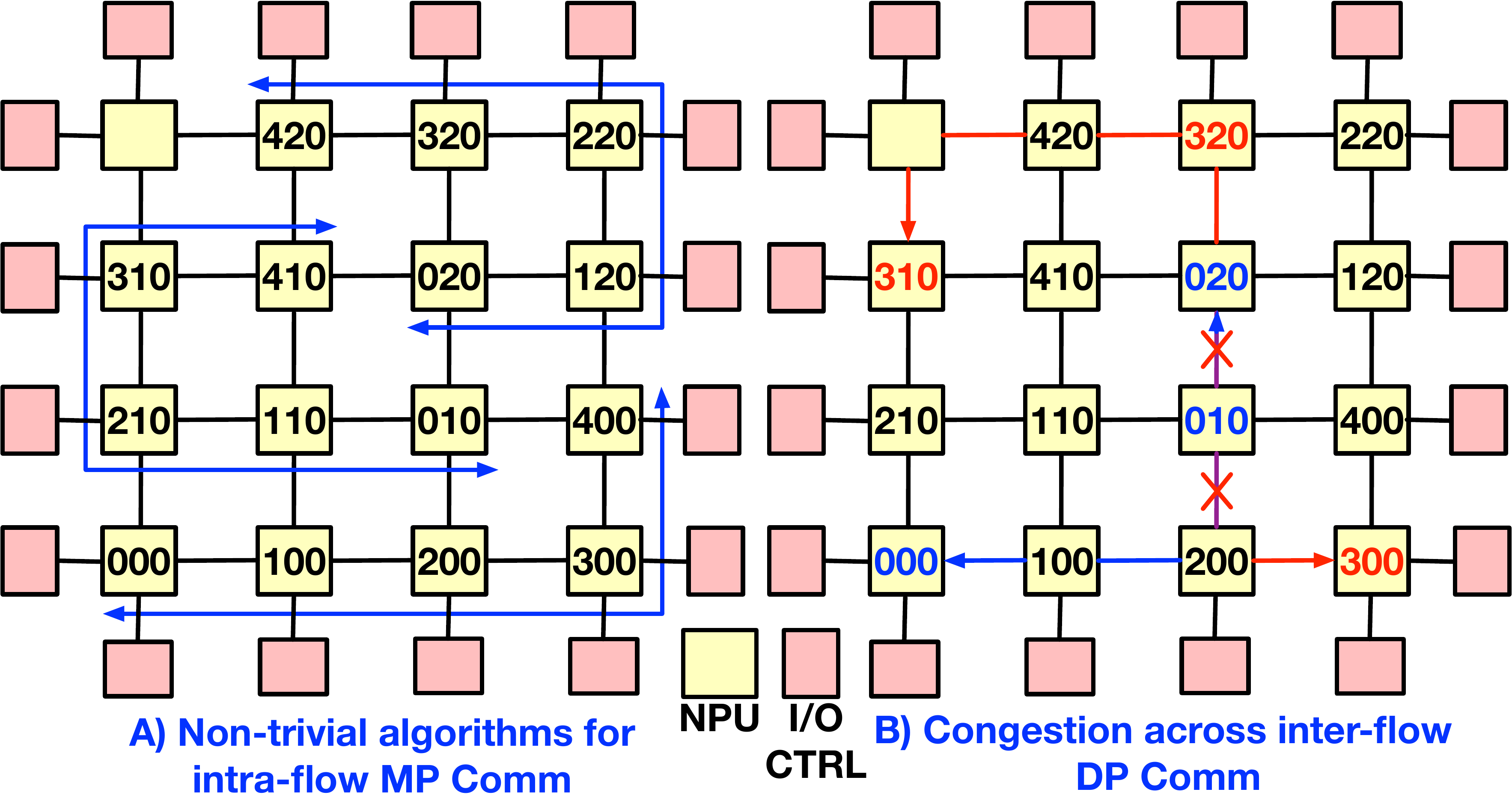}
  \vspace{-1em}
  \caption{\small Network communications on a $4\times 4$ mesh topology for a non-aligned MP(5)-DP(3)-PP(1) parallelization strategy. (A) Non-optimized execution of communication patterns (e.g., All-Reduce) within the NPUs of the same MP group. (B) Traffic congestion between two different DP groups, shown in red and blue, assuming X-Y routing (other group traffic is not shown for clarity).}
  \vspace{-1.3em}
  \label{fig:NetworkCongestion}
\end{figure}

When searching for the best parallelization strategy itself, there are many possible configurations where the size of MP/DP/PP is not aligned with the physical topology dimensions. Such configurations create extra challenges on a 2D Mesh, due to the limited path diversity with distinct NPU-to-NPU distances.

\autoref{fig:NetworkCongestion} illustrates the communication issues within a $4 \times 4$ 2D-mesh topology for an MP(5)-DP(3)-PP(1) strategy. \autoref{fig:NetworkCongestion}(A)~demonstrates how NPUs in the same MP group need to communicate. Collective communications are often optimized for well-structured topologies (e.g., rings, trees, switches). However, as shown in~\autoref{fig:NetworkCongestion}(A), the MP groups form non-standard shapes, making it challenging to identify the most optimized collective algorithm for each shape. For example, the distance between NPU 420 and 020 is two hops, due to the rigid shape of 2D Mesh, \emph{making it impossible to construct a well-constructed ring}, even without considering network congestion. \autoref{fig:NetworkCongestion}(B) depicts the extra traffic congestion between two different DP groups, marked in red and blue, caused by non-aligned dimensions.

\ParaTitle{\ours's Solution.} \ours provides congestion-free topology and routing mechanisms for any size/placement of MP/DP/PP.

\subsubsection{Network BW Utilization}\label{subsub:BWUtilizationChallenge}
Maintaining high bandwidth utilization is challenging for a 2D Mesh. For instance, MP communications are required during both forward-pass and back-propagation phases, while DP communications occur only during back-propagation. However, these links cannot be utilized by MP communications due to the limited paths and lack of optimal routing. Consequently, the links used for DP communication during back-propagation remain underutilized during the forward-pass phase, detrimenting full bandwidth utilization for many strategies on a 2D Mesh.

\ParaTitle{\ours's Solution.} \ours can utilize the full bandwidth of each NPU for every communication phase.

\subsubsection{In-Network Collective Execution}
Supporting in-network collectives can significantly reduce network traffic and improve execution performance as described in~\autoref{subsec:collectivealgs}. This feature, currently employed in off-chip switches~\cite{NVidiaSwitch, MellanoxSHARP}, requires centralized or hierarchical switches which can perform the collection, reduction, and broadcast of multiple data. A 2D Mesh with distributed NPUs and without a shared central entity, however, impedes the adaptation of the in-network collective support.

\ParaTitle{\ours's Solution.} \ours employs a switch-based topology that supports in-network collective execution.

\subsubsection{Takeaway}
Ideally, a fabric for DNN training should enable each NPU to fully utilize its network bandwidth for any communication phase of 3D-parallel training without congestion and with support for in-network collectives. These requirements cannot be met via a 2D Mesh, due to their natural shape and rigidity. This underscores the need for the adaptation of new topology and routing mechanisms, such as \ours.

%% file: content/4_design.tex
\section{\ours\ Network Fabric Architecture}
\label{sec:design}


\begin{figure*}[h]
  \centering
  \includegraphics[width=1\linewidth]{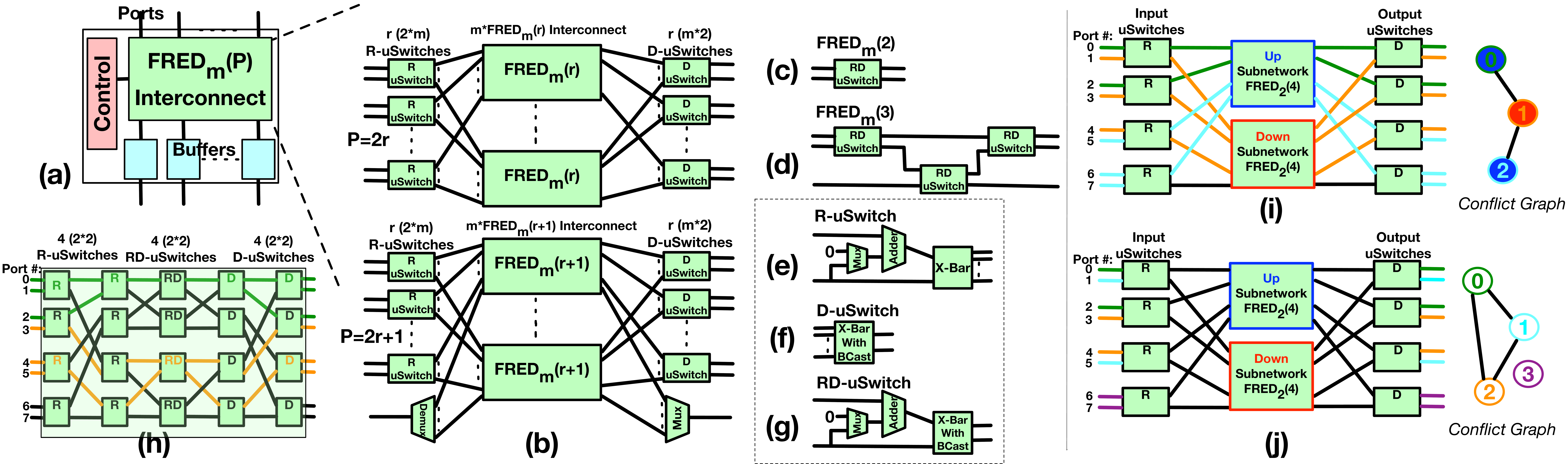}
  \vspace{-2.5em}
  \caption{\small (a) An overview of the \ours switch with P ports. (b) \ours interconnect (recursively constructed) when the number of ports is even ($2r$) or odd ($2r+1$). (c) \oursns$_{m}$($2$) switch. (d) \oursns$_{m}$($3$) switch. (e) R-$\mu$Switch. (f) D-$\mu$Switch. (g) RD-$\mu$Switch. (h) An example of a \oursns$_{2}$($8$) interconnect implementation and two routed All-Reduce communication patterns (green and orange). (i) Routing Algorithm for three All-Reduce comm flows on \oursns$_2$($8$) with conflict graph. (j) Example of Routing conflict.}
  \vspace{-0.5em}
  \label{fig:FredArchitecture}
\end{figure*}

A \ours switch forms the backbone of the fabric.
Hierarchical connections of the \ours switches form the full \ours fabric, which is described in~\autoref{subsec:ScalingFRED}.
The key idea behind a \ours switch is simple: \textbf{break the switch into the most fundamental components, and add small compute capability to each component.}
The fine-grained distribution of compute enables supporting flexible and concurrent in-switch collective execution for 3D parallelism communication patterns.
In addition, distributed computation of collectives is more scalable to map over the high-BW wafer-scale links than having centralized compute and memory entities.

\autoref{fig:FredArchitecture}(a) shows a \ours switch, which consists of a control unit, input port buffers, and the \ours interconnect.
The control unit performs routing between the input ports and the output ports.

The \ours interconnect, shown in~\autoref{fig:FredArchitecture}(b), is inspired by \emph{Clos} networks~\cite{clos}.
Clos networks are identified through the tuple $(m,n,r)$, where $m\geq 2$ is the number of middle stage switches, $n$ is the number of input/output ports per each input/output micro-switch ($\mu$Switch), and $r$ is the number of input/output $\mu$Switches.
\ours's connectivity is similar to the $(m,n=2,r)$ Clos network, which is denoted as \oursns$_m$($P$).
$m$ denotes to the number of middle-stage switches, and $P$ identifies the number of input(output) ports.
\ours can be designed for an arbitrary number of ports by building on top of the previous works \cite{arbitraryBnese}.
$P$ is $\frac{2r}{2r+1}$ when $P$ is an $\frac{\text{even}}{\text{odd}}$ number.
Similar to the Clos network, \ours interconnect is constructed recursively, where the middle stage switches are the $ \frac{m \times \text{\oursns$_m$($r$)}}{\text{$m\times$\oursns$_m$($r+1$)}}$ switches for the $\frac{\text{even}}{\text{odd}}$ number of ports, as shown in~\autoref{fig:FredArchitecture}(b).
The recursive design of \ours ends when encountering the base \oursns$_m$($2$) or \oursns$_m$($3$) Switches, which are depicted in~\autoref{fig:FredArchitecture}(c) and~\autoref{fig:FredArchitecture}(d), respectively.

\textit{\textbf{The main difference of \ours, compared to a baseline Clos, is adding the reduction and/or distribution (broadcast) support}} to the baseline $\mu$Switches.
This creates three types of $\mu$Switches depending on which of these two features is present in the $\mu$Switch.
\autoref{fig:FredArchitecture}(e) shows the \emph{R-$\mu$Switch} structure that has the reduction feature, i.e., reducing data on the two input ports and routing to one of the output ports.
\autoref{fig:FredArchitecture}(f) shows the \emph{D-$\mu$Switch}, which is able to perform distribution by broadcasting one of the input data to both output ports.
\emph{RD-$\mu$Switch} is a $2\times 2$ $\mu$Switch and can perform both reduction and distribution, as shown in~\autoref{fig:FredArchitecture}(g).
The entire \ours switch is built using these three $\mu$Switch types (plus \emph{Muxes} and \emph{Demuxes} to connect the last port to all intermediate stage switches when $P$ is odd) through the recursive process explained earlier.

\autoref{fig:FredArchitecture}(h) shows the complete structure of a \oursns$_{2}$($8$) switch with two concurrent All-Reduce operations (green and orange).
The highlighted $R/D/RD$ means that the reduction/distribution/reduction-distribution features of the corresponding $\mu$Switch are activated.
For instance, the input $\mu$Switch connecting the input ports $4,5$ performs the reduction and routes the result to one of its output ports.
Other non-highlighted $\mu$Switches operate like Clos $\mu$Switches.

%% file: content/5_routing.tex
\vspace{-1em}
\section{Conflict-free Collective Routing}
\label{sec:fred_routing}

\subsection{Communication Patterns on \ours}

The fine-grained reduction and broadcast features enable \ours $\mu$Switches to perform all different types of collective communication patterns observed in distributed training.
Collective implementation on \ours, however, can be abstracted through the notation of \emph{communication flow} (or \textbf{\emph{flow}} in short).

A \emph{flow} on \oursns$_{m}$($P$) includes a set of input ports ($IPs$)=\{ip$_1$, ip$_2$, ...., ip$_i$\} and output ports ($OPs$)=\{op$_1$, op$_2$, ...., op$_j$\}, where $|IPs| \leq P$ and $|OPs| \leq P$.
The \emph{flow} results in reducing the data across the input ports determined in $IPs$ and broadcasting the final result to the output ports identified in $OPs$.
The port numbers and cardinality of $IPs$ and $OPs$ can be set independently, depending on the communication pattern.
Each communication algorithm can be expressed in terms of performing one or more \emph{flows}.
For example, the orange All-Reduce pattern in~\autoref{fig:FredArchitecture}(h) is a single \emph{flow} with $IPs=\{3, 4, 5\}$ and $OPs=\{3, 4, 5\}$.

\ParaTitle{Simple Communication Algorithms.}
Simple communication algorithms refer to communication patterns that can be realized on \ours by performing only one \emph{flow}.
\autoref{table:compoundCollectives} summarizes different simple communication patterns on \ours and the number of involved input/output ports.

\ParaTitle{Compound Communication Algorithms.}
Compound communication algorithms realize the communication patterns through multiple \emph{flows} on \ours.
\autoref{table:compoundCollectives} summarizes different compound communication patterns on \ours.
For example, \emph{Reduce-Scatter} among $i$ inputs is broken into $i$ serial steps of the \emph{reduce} \emph{flow}, and during step $1 \leq j \leq i$, the \emph{reduce} operation corresponding to the result of the $op_j$ is done.
The process is similar for other compound communication algorithms.
\vspace{-2mm}
\subsection{Routing Protocol}\label{subsec:routing}
\ours considers a \emph{flow} as a unit of routing, and supports concurrent routing of multiple \emph{flows}.
Similar to the previous methods \cite{bneseRouting1}, \ours routing protocol is also recursive, meaning that first the status of outermost $\mu$Switch levels (i.e., input/output $\mu$Switches) are determined, and then routing is recursively called on the middle stage switches.
The difference is, however, supporting reduction/distribution features on the \ours $\mu$switches, and the dependency between the input/output ports of a \emph{flow}, which requires a new routing algorithm to realize these differences. \ours's routing protocol is built upon the following intuitions:

\begin{itemize}
    \item If two flows share the same input or output $\mu$Switch, they should be routed through different middle-stage switches (subnetworks).
    \item If both input ports of an R-$\mu$Switch or RD-$\mu$Switch belong to the same \emph{flow}, the reduction feature is activated.
    \item If both output ports of a D-$\mu$Switch or RD-$\mu$Switch belong to the same \emph{flow}, the distribution (broadcast) feature of the $\mu$Switch is activated.
\end{itemize}

The latter two points are easy to realize.
To satisfy the first point, \ours routing protocol creates a \emph{conflict graph}.
\autoref{fig:FredArchitecture}(i) shows the first step of a routing example for a \oursns$_2$($8$) interconnect with the associated conflict graph for this step.

In the conflict graph, each node represents a \emph{flow} and the edges between the nodes represent a conflict (i.e., sharing an input or output $\mu$Switch) between the two nodes (\emph{flows}).
\ours routing applies the graph coloring on the conflict graph to find the routing of each \emph{flow}.
The number of colors is the number of intermediate stage switches (i.e., $m$).
\autoref{fig:FredArchitecture}(i) also shows the results of the graph coloring.
Here, there are only two colors since $m=2$.
Two flows are routed to the up subnetwork (blue), and one to the down subnetwork (red).
After this step, the routing protocol and the conflict graph generation are recursively called on the middle blue and red \oursns$_2$($4$) switches.
Note that a desired property of DL training is the deterministic and repetitive nature of its communication patterns that can be inferred at compile time.
Therefore, the routing algorithm for different comm phases of the training workload can be executed at compile time and then saved at the control unit of the \ours switches and used during the training to minimize the routing overhead.

\subsection{Routing Conflicts and Methods to Resolve}\label{subsec:routingConflict}

There are certain cases where not all \emph{flows} can be routed at the same time, causing \emph{routing conflict}.
The routing conflict is identified when the graph coloring fails to color all of the nodes within the conflict graph.
\autoref{fig:FredArchitecture}(j) shows an example of a routing conflict when there are four \emph{flows} to be routed on a \oursns$_2$($8$) and the resulting conflict graph.
The conflict graph cannot be colored using only two colors due to the circular dependencies between \emph{flows: 0, 1, 2}.
Note that the routing conflict may happen during any recursive call to the routing algorithm (for routing the subnetworks).
If the routing conflict is identified, the entire routing is marked to have a conflict.


\input{tables/compoundCollectives}
We now discuss ways to address such conflicts.

 \ParaTitle{(1) Blocking the Conflicting \emph{Flows}.}
The first trivial way is to block some of the conflicting \emph{flows} and run them after the other \emph{flows} are finished.
This translates to removing some of the nodes in the conflict graph.
For example, in~\autoref{fig:FredArchitecture}(j), if any of the \emph{flows} $1, 2,$ or $3$ is blocked, the routing can proceed to the next step (i.e., subnetworks).
This option is, however, costly in terms of performance since it blocks some of the flows.

\ParaTitle{(2) Increasing the Number of Middle Stages.}
Another method is to design \ours switches with more intermediate stage switches (i.e., increase $m$).
This method increases the number of colors for the graph coloring algorithm.
Therefore, more conflicting \emph{flows} can be routed simultaneously.\footnote{For example, \oursns$_3$($8$) can route all the flows in~\autoref{fig:FredArchitecture}(j).}
However, this comes at the expense of more HW overhead.

\ParaTitle{(3) Decomposing the Communication Algorithms.}
For the unicast-only traffic, \ours interconnect is \emph{rearrangeably nonblocking} when $m=2$ and \emph{strict-sense nonblocking} when $m\geq 3$.
This fact can be leveraged to decompose some of the communication algorithms into multiple steps and break the dependency among input/output ports in each step (i.e., making them unicast traffic).
In the worst case, any collective algorithm can be decomposed into complete unicast traffic.
For example, All-Reduce can be handled through a ring-based algorithm at the endpoints (NPUs), rather than in-network execution, which is complete unicast traffic.
As a result, \emph{flows} $0, 1,$ and $2$ in~\autoref{fig:FredArchitecture}(j) can switch to ring-based All-Reduce at the endpoint, while \emph{flow} $3$ uses an in-network All-Reduce algorithm.
This method solves the routing by degrading the communication performance of the conflicting \emph{flows} (but it does not block any \emph{flow}).

\ParaTitle{(4) Intelligent Device Placement.}\label{subsec:fredDevicePlacement}
Another method to prevent conflicts is through intelligent device placement (mapping) of the training workers to the physical NPUs at the start time.
For example, if in~\autoref{fig:FredArchitecture}(j) the workers mapped to NPUs of ports $1$ and $4$ swap their locations, the conflict does not happen.

\emph{In \ours, we prioritize the communication performance and do not use options (1) and (3).
We use option (2) to simplify the device placement algorithm by only using \oursns$_{3}$($P$) switches, ensuring that we have three colors in our routing algorithm protocol.
Then, for the device placement algorithm, we map the training workers within the same MP group on consecutive physical NPUs, followed by iterating over workers within PP and DP, respectively.
This is sufficient to prevent routing conflicts for 3D-Parallelism communication patterns.}

\subsection{Handling Overlapping Communications}\label{subsec:overlappedcomms}
In training, the workload at a given time may require multiple communication operations. For example, while handling the DP communication in backpropagation, the workload may initiate the PP communication to exchange the next microbatch between the workers. However, FRED's circuit switch configuration may handle one communication phase at a given time. Additionally, different NPUs might issue communication at different times, due to variations in the compute latencies. Hence, there should be a mechanism to safely preempt the current executing communication operation and execute the new communication, with minimal effects to the in-flight packets, if the latter has a higher priority.


We address this issue by allocating multiple Virtual Circuits (VCs) per port, each dedicated to a specific communication group (e.g., MP), and the FRED's interconnect to be reconfigured between different overlapping communication operations. While it is possible to frequently reconfigure FRED's interconnect in short intervals to handle overlapping communication operations concurrently, we choose to reconfigure FRED to execute the highest priority communication operation among the currently pending operations (and preempt the current communication if a new higher priority communication is issued). This decision simplifies the design and minimizes the FRED's reconfiguration overhead, and is in line with the training workload requirements, since the workload is usually blocking on one communication operation (highest priority) at any given point in time. In our 3D-parallel case, the priority of communication operations in descending order is: MP, PP, and DP. More discussion on FRED's buffer management and flow control is described in \autoref{subsub:fredTopology}.

%% file: tables/compoundCollectives.tex
\begin{table}[h]
\small
\caption{\small Simple (shaded) and Compound collective algorithms.}
\label{table:compoundCollectives}
\vspace{-1em}

\resizebox{0.80\columnwidth}{!}{
\begin{tabular}{cccl}
\hline
\textbf{Pattern} & \textbf{\textit{|IPs|}} & \textbf{\textit{|OPs|}} & \textbf{Comment} \\
\hline
\rowcolor{lightgray} Unicast & 1 & 1 & \\ \hline
\rowcolor{lightgray} Multicast & 1 & \textgreater{}1 & \\ \hline
\rowcolor{lightgray} Reduce & \textgreater{}1 & 1 & \\ \hline
\rowcolor{lightgray} All-Reduce & i\textgreater{}1 & i\textgreater{}1 & \begin{tabular}[c]{@{}l@{}}Input ports and output\\ ports are the same\end{tabular} \\ \hline
Reduce-Scatter & i\textgreater{}1 & i\textgreater{}1 & \begin{tabular}[c]{@{}l@{}}
Broken into multiple serial \\ Reduce collectives, each \\ on a different output port\end{tabular} \\ \hline
All-Gather & i\textgreater{}1 & i\textgreater{}1 & \begin{tabular}[c]{@{}l@{}}Broken into multiple serial \\ Multicast collectives, each \\ on a different input port\end{tabular} \\ \hline
Scatter & 1 & i\textgreater{}1 & \begin{tabular}[c]{@{}l@{}}Broken into multiple serial \\ Unicast operations, each \\ on a different output port\end{tabular} \\ \hline
Gather & i\textgreater{}1 & 1 & \begin{tabular}[c]{@{}l@{}}Broken into multiple serial \\ Unicast operations, each \\ on a different input port\end{tabular} \\ \hline
All-To-All & i\textgreater{}1 & i\textgreater{}1 & \begin{tabular}[c]{@{}l@{}}Broken into i serial steps of \\ Unicast operations. In step \\ 1 $\leq$ j $\leq$ i, each input port \\ unicasts to the output port \\ with distance j in the OPs\end{tabular} \\
\bottomrule
\end{tabular}
}

\vspace{-1em}
\end{table}

%% file: content/6_architecture.tex
\section{Wafer Scale Architecture}\label{sec:methodology}

We present an instance of a wafer-scale NPU system connected using \ours, for evaluation purposes.
We note that alternate configurations are also feasible.

\subsection{Layout of \ours\ Fabric}\label{subsec:ScalingFRED}

\begin{figure}[h]
  \centering
  \includegraphics[width=1\linewidth]{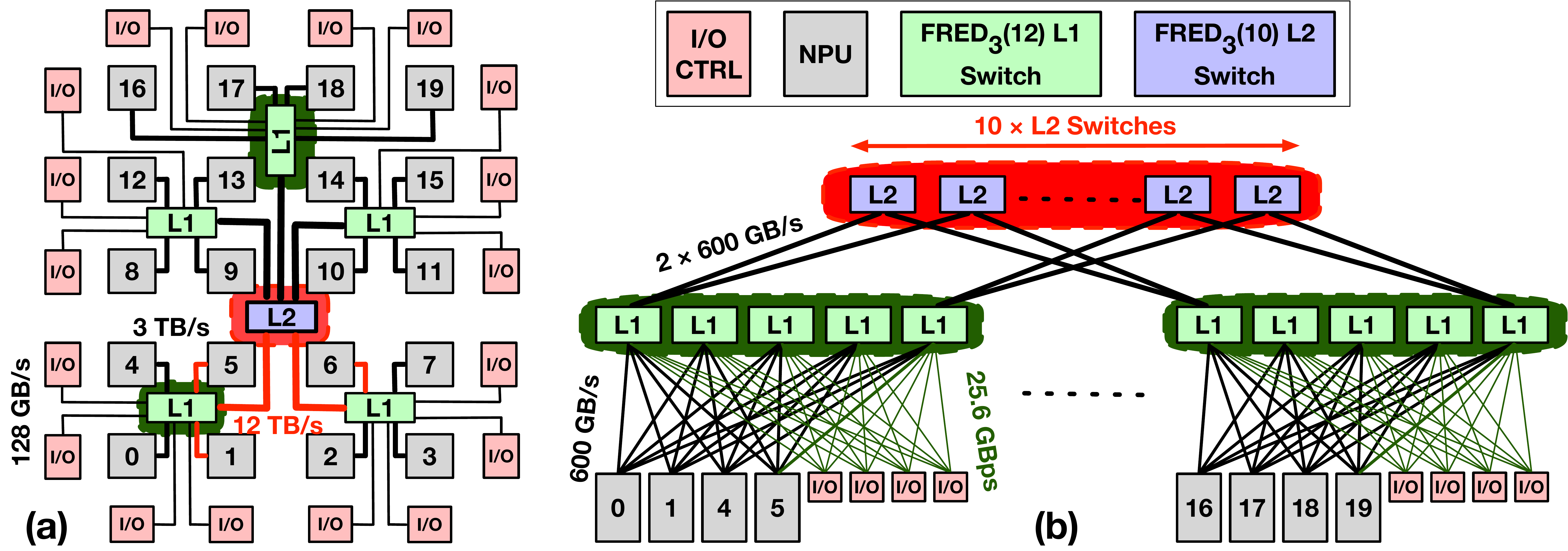}
  \vspace{-2em}
  \caption{\small Physical and Logical Views of 2-level \ours\ Topologies.}
  \vspace{-1em}
  \label{fig:FredLayoutPhysicalLogical}
\end{figure}

A \ours\ switch builds a foundation to connect multiple wafer-scale NPUs.
However, for large wafer-scale systems, due to physical limitations such as wiring, area, etc., it is not feasible to connect all of the NPUs through a single \ours\ switch.
Hence, the \emph{\ours\ fabric} provides a hierarchical design for the scalable connection of large wafer-scale systems.
\autoref{fig:FredLayoutPhysicalLogical} shows an example of the \ours\ fabric that shows a 2-level tree connection of the \ours\ switches and the NPUs connected to the leaf (\emph{L1}) switches\footnote{We note that \ours\ layout shown is not tiled. This means that the substrate (where chiplets are bonded) may not be able to use stepper-based lithography. But direct-written maskless lithography is not uncommon for substrate patterning. This was used in a commercial packaging provider ThinkDeca~\cite{masklessUsage}. Such patterning has no symmetry requirement, albeit it has a lower throughput. Also, note that using maskless lithography increases the substrate manufacturing moderately~\cite{masklessLithoCost}, but substrate manufacturing is a small fraction of the total system cost~\cite{WaferScaleChiplets}.}.
In general, tree height and the BW across different levels are determined by the system size and physical constraints (see~\autoref{subsec:systemConfig}).

When there are multiple levels of \ours\ switches, the communication algorithms might need to cross several switches and hence, need to be optimized accordingly.
For example, \autoref{fig:FredLayoutPhysicalLogical}(a) shows the flow path for an All-Reduce between NPUs $1, 5,$ and $6$.
In this case, the data of NPUs $1\text{ and }5$ are reduced on their local L1 switch (to reduce the traffic going to the L2 switch), and the result along with the data of NPU $6$ are reduced on the L2 switch.
The final result is sent back to the corresponding L1 switches.
The L1 switch attached to NPUs $1\text{ and }5$ also multicasts the result to the NPUs.


\subsection{Wafer-scale Architecture Configuration}\label{subsec:systemConfig}

We assume a standard 300 $mm$ wafer diameter, similar to the prior works~\cite{waferScaleGPU,packageLessProcessing}, resulting in a 70000 $mm^2$ wafer area. 

\subsubsection{Constraints}
Fundamentally, there are two physical limitations that limit the amount of compute and other resources on the wafer: (i) Thermal constraints, and (ii) Power delivery network~\cite{WaferScaleChiplets,waferScaleGPU,packageLessProcessing}. 
Thermal constraints limit the amount of power that can be delivered to the wafer, depending on the cooling mechanism. 
Previous works report the maximum power limit within the $9.6\:KW$~\cite{waferScaleGPU} to  $15\:kW$~\cite{Cerebras} range. 
In this paper, we assume $\pmb{15\:kW}$ power is available for the wafer-scale system.
The other limitation is the power delivery network, which might necessitate using big on-wafer \emph{voltage regulator modules (VRMs)}, limiting the available area for NPUs~\cite{waferScaleGPU}. 
However, alternative solutions can eliminate the need for on-wafer VRMs by either supplying the voltage from the top of the wafer~\cite{Cerebras}, or delivering the power from the back of the wafer by using the \emph{through-wafer-vias (TWVs)}~\cite{TWV}. 
In this paper, we assume the \textbf{on-wafer VRMs are not used} by using any of the solutions described earlier. 

\subsubsection{Physical System Parameters}
\label{sec:physical_param}
\autoref{tab:config} shows the other set of physical parameters.
We assume that the NPU chiplets are tested before bonding.
If Known Good Die testing is difficult, larger chiplets such as NPU Compute may need to be broken into smaller constituents.
Recent work~\cite{ChipletSmall} has suggested that these chiplets actually need to be moderately large (40$mm^2$-400$mm^2$) in size for cost-optimality.
For the purposes of our evaluation, we assume an H100 GPU-like NPU compute chiplet, each equipped with five stacks of HBM3 chiplet memories, resulting in combined power consumption of $700\:W$ and an area of $1314\:mm^2$~\cite{NvidiaH100}. 

The NPU compute chiplet perimeter can support up to 12 TBps wafer-scale BW, where $6$ TBps of it is allocated to support the 3 TBs local HBM memory BW ($3$ TBps for read + $3$ TBps for write), and the other $6$ TBps is allocated to support $3$ TBps bi-directional total NPU-to-NPU BW ($3$ TBps for send + $3$ TBps for receive). 

The $15\:KW$ power budget limits the total amount of NPUs on the wafer to $15\:KW/700\:W\approx21$, excluding other component power overheads (e.g., I/O controller, wafer-scale wires).
This anticipated power density of 22W/cm$^2$ is well within the projection of cooling capability in heterogeneous integration roadmaps~\cite{HIR}.
In this paper, we consider a $20$-NPU wafer-scale system to make room for other component power overheads.
Additionally, $18\times$I/O Controllers are used to connect the wafer to the external memory.
Hence, the total NPU $+$ I/O Controller area overhead is $26640\:mm^2$. 

Similar to~\cite{waferScaleGPU}, we assume in the baseline, the NPU chips are placed with a 100 $um$ distance from each other. Combined with the I/O controllers, the entire baseline can be fit within a rectangle with the size of 190.8 $mm$ $\times$ 150.4 $mm$ in the center of the wafer, leaving the rest of the wafer area unclaimed.

\input{tables/config}
\input{tables/overhead}

\subsubsection{\ours\ Topology and Parameters}\label{subsub:fredTopology}
To motivate \ours, we leverage the fact that the combination of a constrained power budget and high-end NPUs results in utilizing $26640\:mm^2$ out of $70000\:mm^2$ area, \textbf{making room to utilize otherwise unclaimed area for flexible fabrics like \ours}. However, any fabric proposal must have low power consumption since most of the power budget is allocated to the NPUs. \textbf{We demonstrate that \ours\ meets these properties.}

Our target \ours\ topology is similar to~\autoref{fig:FredLayoutPhysicalLogical}(a), where $20$ NPUs and I/O controllers are connected through a 2-level (almost) fat-tree topology. Similar to the baseline, the BW/NPU is still $3$ TBps, but the bisection BW is increased to $30$ TBps. It is almost fat-tree since the L1-to-L2 BW is the summation of attached NPU BW only (and not NPU $+$ I/O Controller). The reason is that if one participant of any \emph{flow} (e.g., \emph{Reduce}) is an I/O controller, then the entire \emph{flow's} BW requirement is determined by the I/O controller's BW (e.g., 128 GBps), which is significantly less than NPU-to-NPU BW. Hence, an almost fat-tree gives the same performance as the full fat-tree.

Looking at the BW requirements of \ours\ L1/L2 switches in~\autoref{fig:FredLayoutPhysicalLogical}(a), it is clear that each switch chiplet requires a perimeter, to connect the wafer-scale network wires, that is not feasible to build. Hence, in reality, each of the \ours\ switches in~\autoref{fig:FredLayoutPhysicalLogical}(a) is decomposed into multiple lower-BW \ours\ chiplets. \autoref{fig:FredLayoutPhysicalLogical}(b) shows a logical view of implementing the (almost) fat-tree based topology of~\autoref{fig:FredLayoutPhysicalLogical}(a) using feasible \ours\ chiplets. As~\autoref{fig:FredLayoutPhysicalLogical}(b) shows, each switch of~\autoref{fig:FredLayoutPhysicalLogical}(a) is implemented by decomposing it into multiple smaller, but feasible, \ours\ switches (enclosed in the strip line). For our evaluations, we use \oursns$_3$($P$) switches. 

As~\autoref{fig:FredLayoutPhysicalLogical}(b) shows, in \ours\ fabric, L1 switches have hybrid BW downstream links to connect to the NPUs and I/O controllers. This requires \ours\ L1 switches to use different interface circuitry for NPU vs. I/O controller links, which is accounted for in the overhead numbers in~\autoref{tab:overhead}. In general, hybrid on-chip interconnects are widely used in many designs (e.g., to connect on-chip routers vs. memory controllers in multi-core processors)~\cite{onchipbook}.

\ParaTitle{Flow Control.}
We assume a Virtual Cut-Through flow control with a credit-based backpressure mechanism to guarantee the switch buffer as packets flow through FRED's fabric. To enable preemptive communication execution, we consider four VCs per port: three data VCs dedicated to MP, DP, and PP packets and one control VC for the ACK/NACK and other control messages. The data/control packet size is 4KB/512B, with each flit size set to be 512B. The packet header size is 6B to allow for large sequence numbers. Each packet header also has the index to the $\mu$Switch configuration bits, stored in the control unit for a specific communication phase\footnote{Compound collectives have multiple phases}. If all ports receive a packet belonging to a higher priority phase,  \ours changes its $\mu$Switch configuration to that phase and starts forwarding the packets from that phase. Additionally, there is a default header index, which refers to a phase where all flows are unicast and \ours falls back to the online routing to determine the $\mu$Switch configs. While not present in our workloads, this mode is useful when dealing with communication patterns such as \textit{alltoallv} where different src/dst pairs have different size flows that are changing dynamically. 

The retransmission protocol is set to be simple Go-Back-N, with an accumulative ack per every 16 data packets to reduce the ack overhead to less than $1\%$ of the network BW. If a switch receives a NACK from an NPU, it forwards it to all input ports participating in that flow, which is then propagated to all NPUs serving as the source of the flow, and retransmission starts from the NACKed packet.


Additionally, each input port has a 24KB buffer per data VC and a 2KB buffer for the control VC. These policies ensure that in the case of communication preemption, there are enough buffers available (i.e., $link\_BW\times RTT=\text{24KB}$) for the new communication operation to send at the full link BW.

\ParaTitle{HW Overhead.}
\autoref{tab:overhead} shows the overheads of our proposed \ours implementation shown in~\autoref{fig:FredLayoutPhysicalLogical}. We assume 1.5KB SRAM per FRED switch to store the $\mu$Switch configurations for different communication operations. The numbers are obtained post layout using 15nm NanGate PDK. The total power overhead is $179.35\:W$, which is about $1.2\%$ of the total power budget. The total area overhead is $25195\:mm^2$, which can be accommodated by using the unclaimed area available on the wafer. Note that, as discussed in~\autoref{subsub:fredTopology}, the main area overhead of the \ours\ chiplets is due to I/O for supporting high-BW wafer-scale interconnects, and not because of the switch logic overhead.

\ParaTitle{Discussion: \ours\ Area Overhead.}
As we discussed earlier, the unclaimed area on the wafer allows for designing large (but low power) \ours\ switches to deliver high I/O BW requirements for our topology. In fact, \ours's internal logic occupies less than 5\% of the chip area. Hence, the area overhead of \ours\ can be significantly reduced if the I/O density increases.

\input{tables/configs.tex}

In our design, we conservatively assume the switch chips use the same interconnect technology as the NPUs (e.g., pitch, frequency, etc.). However, switch area can be further reduced by applying more aggressive network bandwidth technologies. Next generation of I/O technology is expected to deliver up to 250 GBps/mm (compared to 107.4 GBps/mm in our design)~\cite{waferBWDensity1}. This results in designing \ours\ switch chips with only 18.4\% of current area with the same I/O BW.

The other I/O technology alternative is using the serialized high-speed links such as UCIe Advanced~\cite{waferBWDensity2}, which can deliver up to 1 TBps/$mm$. This results in designing \ours\ switch chips with only 5\% of the current area. Note that even with the high area assumption of \ours, we don't expect the yield issue to be a practical problem since compared to the compute NPUs, \ours\ switches have much less internal logic and hence encounter fewer defects.

%% file: tables/config.tex
\begin{table}
\small
\caption{\small Physical system parameters.}
\label{tab:config}

\resizebox{0.9\columnwidth}{!}{
\begin{tabular}{lllll}
\toprule
\textbf{Component} & \textbf{Model} & \textbf{Area} & \textbf{Power} & \textbf{Characteristics} \\
\midrule
\begin{tabular}[c]{@{}l@{}}NPU\\ Compute\end{tabular} & \begin{tabular}[c]{@{}l@{}}GPU-like\\ \cite{NvidiaH100}\end{tabular} & 814 $mm^2$ & 525 $W$ & \begin{tabular}[c]{@{}l@{}}• FP16:\\ 1,000 TFLOPS\end{tabular} \\ \hline
\begin{tabular}[c]{@{}l@{}}NPU\\ Memory\end{tabular} & \begin{tabular}[c]{@{}l@{}}5 $\times$\\ HBM3\\ \cite{NvidiaH100,waferScaleGPU} \end{tabular} & \begin{tabular}[c]{@{}l@{}}5 $\times$\\ 100 $mm^2$\end{tabular} & \begin{tabular}[c]{@{}l@{}}5 $\times$\\ 35 $W$ \end{tabular}& \begin{tabular}[c]{@{}l@{}}• Total Capacity:\\ 80 GB\\ • Total BW:\\ 3 TBps\end{tabular} \\ \hline
\begin{tabular}[c]{@{}l@{}}Wafer-Scale\\ Interconenct\end{tabular} & \begin{tabular}[c]{@{}l@{}}SI-IF\\ \cite{waferScaleGPU,packageLessProcessing} \end{tabular} & \begin{tabular}[c]{@{}l@{}}4 $um$\\ pitch\end{tabular} & \begin{tabular}[c]{@{}l@{}}0.063\\ pJ/bit~\cite{pal2020pathfinding}\\ Baseline:\\ $100\:W$\end{tabular} & \begin{tabular}[c]{@{}l@{}}• \# of \\Metal Layers: 2\\ • Freq: 1.74 GHz\\ • BW: 53.7 GB\\/mm/metal layer\\ • BW/NPU\\ Compute: \\ 6 TBps\\/metal layer\\ Latency: 20 $ns$\\\end{tabular} \\ \hline
\begin{tabular}[c]{@{}l@{}}I/O\\ Controller\end{tabular} &\begin{tabular}[c]{@{}l@{}}CXL 3\\ \cite{cxl3}\end{tabular} & \begin{tabular}[c]{@{}l@{}}18 $\times$\\ 20 $mm^2$\end{tabular} & \begin{tabular}[c]{@{}l@{}}18 $\times$\\ 5 $W$\end{tabular} & \begin{tabular}[c]{@{}l@{}}• BW: 128 \\GBps/controller\end{tabular}
\\
\bottomrule
\end{tabular}
}

\vspace{-0.5em}
\end{table}

%% file: tables/overhead.tex
\begin{table}
\small
\caption{\small HW overhead of \ours implementation of~\autoref{fig:FredLayoutPhysicalLogical}(b).}
\label{tab:overhead}
\vspace{-1em}

\resizebox{0.9\columnwidth}{!}{
\begin{tabular}{crr}
\toprule
\textbf{Component} & \textbf{Area} & \textbf{Power} \\
\midrule
\begin{tabular}[c]{@{}l@{}}\oursns$_{3}$($12$) L1 Switch\end{tabular} & $15 \times 685 \: mm^2$ & $15 \times$ 3.75 $W$ \\
\begin{tabular}[c]{@{}l@{}}\oursns$_{3}$($11$) L1 Switch\end{tabular} & $10 \times 678 \: mm^2$ & $10 \times$ 3.40 $W$ \\
\begin{tabular}[c]{@{}l@{}}\oursns$_{3}$($10$) L2 Switch\end{tabular} & $10 \times 814 \: mm^2$ & $10
\times$ 3.11 $W$ \\
\begin{tabular}[c]{@{}l@{}}Additional Wafer-Scale Wiring\end{tabular} & N/A & $58\:W$ \\
Total & $25195 \: mm^2$ & 179.35 $W$ \\
\bottomrule
\end{tabular}
}

\end{table}

%% file: tables/configs.tex
\begin{table}
\small
\caption{\small Target configurations.}
\label{fred:tab:configs}
\vspace{-1em}

\resizebox{0.9\columnwidth}{!}{
\begin{tabular}{cl}
\toprule
\textbf{Configs} & \textbf{Comment} \\
\midrule
Baseline & \begin{tabular}[c]{@{}l@{}}The baseline topology where 20 NPUs are formed using a \\4$\times$5 2D-mesh topology (3.75 TBps bisection). Each border \\NPU is attached to an I/O controller, except for the corner \\NPUs that are connected to two I/O controllers, summing \\the total I/O controllers to 18.\end{tabular} \\ \hline
\ours-A & \begin{tabular}[c]{@{}l@{}}\ours topology where 20 NPUs and 18 I/O controllers are \\connected similar to \autoref{fig:FredLayoutPhysicalLogical}, but with the \\\textbf{same bisection BW as the baseline (3.75 TBps)}. \\This is achieved by downscaling L1-L2 links from \\$12$ TBps, in \autoref{fig:FredLayoutPhysicalLogical}(a), to $1.5$ TBps. Additionally, it \\\textbf{does not support in-switch collective execution.}\\ Thus, collective algorithms are handled at the endpoints.\end{tabular} \\ \hline
\ours-B & \begin{tabular}[c]{@{}l@{}} Similar to \ours-A, \textbf{but with in-network collective} \\\textbf{execution.}\end{tabular} \\ \hline
\ours-C & \begin{tabular}[c]{@{}l@{}}\ours topology where 20 NPUs and 18 I/O \\controllers are connected similar to \autoref{fig:FredLayoutPhysicalLogical} \\\textbf{(30 TBps bisection),} but \textbf{without} any in-switch \\collective execution (handled at the endpoint). \end{tabular} \\ \hline
\ours-D & \begin{tabular}[c]{@{}l@{}}Similar to \ours-C, but \textbf{with in-network}\\ \textbf{collective execution.}\end{tabular} \\
\bottomrule
\end{tabular}
}

\vspace{-1em}
\end{table}

%% file: content/7_methodology.tex
\input{tables/workloads}

\section{Evaluation Methodology}

\subsection{Baseline and \ours Configurations}

\ParaTitle{Baseline.} The baseline topology is a $5\times4$ 2D-mesh with I/O controllers attached to the edge NPUs, similar to prior multi-chiplet wafer-scale prototypes~\cite{waferScaleGPU, WaferScaleChiplets,Simba,tto,chipletcloud}.
Since each NPU has 3 TBps bandwidth (\autoref{sec:physical_param}), 
each NPU-to-NPU link in the 2D-Mesh is equal to $750$ GBps, resulting in the bisection BW of 3.75 TBps.
The I/O Controller-to-NPU is $128$ GBps.

\ParaTitle{\ours.} We test four different variations of \ours to show how different features of \ours contribute to the overall performance.
\autoref{fred:tab:configs} shows the target configurations.
\emph{\ours-A} shows the effect of going from mesh to switch-based topology with the same bisection and without in-network collective execution.
\emph{\ours-B} builds on top of \ours-A and adds the in-network collective execution feature.
\emph{\ours-C} increases the bisection BW without in-network collective execution.
Finally, \ours-D is the most optimal variant of \ours by adding the in-network collective execution to the previous variant.

\vspace{-2.8mm}
\subsection{Collective Algorithm}
For the baseline 2D mesh and when there is a wafer-wide collective, we use the hierarchical 2D algorithm with two concurrent chunks (in reverse direction) to enhance utilization~\cite{mesh_allreduce,tto}.
For collectives between arbitrary NPUs, we build logical rings between involved NPUs and perform the ring algorithm.
We also use X-Y routing, which is common in real systems \cite{mesh_allreduce}.
For \ours-A and \ours-C, we use the hierarchical 2-D ring algorithm to reduce the traffic of L1-L2 links, similar to \cite{blueconnect}.
\ours-B and \ours-D use the in-network capability and use the hierarchical \ours switch topology to perform the collective, as explained in~\autoref{subsec:ScalingFRED}.

\subsection{Target Workloads and Execution modes}\label{subsec:targetWorkloads}
In the interest of space, we evaluate four training workloads, ranging from 60M to 1T parameters to be the representative for a broad range of ML workloads. \autoref{tab:workloads} shows the target workloads and their corresponding parallelization strategy and execution models studied in~\autoref{subsec:result_deepdive}.
\resnet and \transsmall (Transformer model with 17 billion parameters) can fit on the on-wafer memory and hence, use the \emph{weight stationary} execution mode (\autoref{sec:execution_modes}).
In contrast, \gpt and \translarge (Transformer model with 1 trillion parameters) use the \emph{weight streaming} execution mode (\autoref{sec:execution_modes}).
Workers within the same DP group perform All-Reduce together during the back-propagation to sync on weight gradients. In \emph{weight stationary} mode, the workers use the Microsoft ZeRO optimizer stage 2 \cite{mszero} along the DP dimension to reduce the memory footprint.
Note that in \emph{weight streaming} mode, the DP groups should reduce the gradients as they stream them out to the external memory through the I/O controller.
The pattern is the reverse communication direction of~\autoref{fig:DataIngestion}. For \transsmall, \gpt and \translarge, the model split is based on the Megatron-LM method \cite{megatronlm}, which requires two All-Reduces (along the MP dimension) for each transformer layer stack during forward-pass \& back-propagation.
For the PP split on \transsmall, we assume the minibatch is divided into 8 microbatches to hide the effect of pipeline bubbles \cite{GPipe}.
For \gpt, however, pipelining works differently since it is combined with the weight streaming.
In this case, $PP\:=\:2$ indicates that each time $2$ consecutive layers are brought to the wafer and distributed among different NPUs along the PP dimension.
Thus, splitting the minibatch into two microbatches is enough to hide the pipeline latency.
In \autoref{subsec:microbench} and \autoref{subsec:result_deepdive}, the minibatch size for all workloads is set to DP\_size$\times 16$, while in \autoref{subsec:result_various} (and also~\autoref{fig:Motiv}) the minibatch size is increased to DP\_size$\times 40$ to allow for finer-grain pipelining when PP\_size increases\footnote{For these results, we assume the number of microbatches is 1, 10, 20, 20, 20, 40 for the \transsmall with $PP$ size of 1, 2, 4, 5, 10, 20, respectively. For \translarge, the number of microbatches is equal to the $PP$ size.}. All workloads use FP16 gradient precision. 

\subsection{Simulation Framework}\label{subsec:simFramework}

We use ASTRA-SIM~\cite{astrasim,AstraSimGithub}, which is an open-source simulation methodology for modeling distributed training systems. ASTRA-SIM enables the profiling of compute and communication performance of distinct wafer-scale fabrics, including \ours.
It can model various parallelization strategies and the overlapping of compute with comm kernels.
Additionally, its network back-end can simulate the comm operations in detail.
We extend ASTRA-SIM to model the I/O-to-wafer transfers for both the weight stationary and weight streaming scenarios.
For each workload, we run the simulation for two training iterations (i.e., two forward + two backward-pass).

Previous works have shown that endpoint-based collective execution (our baseline) puts more pressure on the endpoint's compute and memory BW resources, hindering the compute kernel efficiency \cite{saeedACE}. To favor the baseline and only focus on the network characteristics, we omit such effects in our baseline system and assume the compute kernels can run as efficient as the in-network collective execution systems such as \ours.

\ParaTitle{Metric of Evaluation.}
In~\autoref{sec:workloadResults}, we report the end-to-end training times and their breakdowns into total compute time and different \emph{exposed} communication times. Since the minibatch size per training iteration may be different depending on the parallelization strategy, we normalize the reported times by dividing the latencies by the minibatch size when comparing the different parallelization strategies of the same workload (e.g., \autoref{fig:Motiv}). 
The exposed communication time refers to the amount of time that is not overlapped with the compute time and the workload is waiting for the communication to be finished.
Depending on the parallelization strategy and execution model, there might be multiple sources of exposed communication times—load, DP, MP, PP, and/or weight streaming.

%% file: tables/workloads.tex
\begin{table}
\small
\caption{\small Target workloads.}
\label{tab:workloads}
\vspace{-1em}

\resizebox{0.9\columnwidth}{!}{
\begin{tabular}{clc}
\toprule
\textbf{Workload} & \begin{tabular}[c]{@{}l@{}}\textbf{Parallelization}\\ \textbf{Strategy}\end{tabular} & \textbf{Execution Model} \\
\midrule
ResNet-152 \cite{ResNet} & MP(1)\_DP(20)\_PP(1) & \emph{Weight Stationary} \\
Transformer-17B \cite{Transformer17B} & MP(3)\_DP(3)\_PP(2) & \emph{Weight Stationary} \\
GPT-3 \cite{GPT3} & MP(2)\_DP(5)\_PP(2) & \emph{Weight Streaming} \\
Transformer-1T \cite{Transformer1T} & MP(1)\_DP(20)\_PP(1) & \emph{Weight Streaming} \\
\bottomrule
\end{tabular}%
}

\vspace{-1em}
\end{table}

%% file: content/8_results.tex
\section{Results}
\label{sec:results}
\label{sec:workloadResults}

\subsection{Microbenchmark Results.} \label{subsec:microbench}
\begin{figure}[h]
  \centering
  \includegraphics[width=0.8\linewidth]{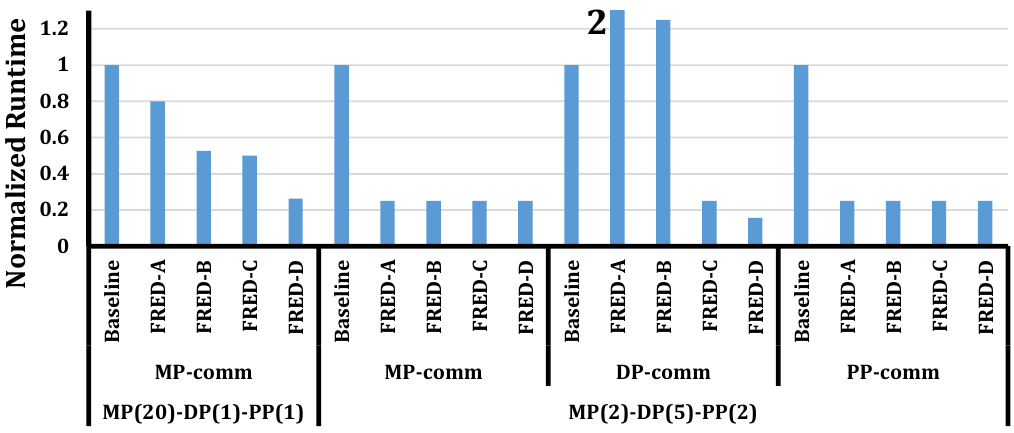}
  \vspace{-1.5em}
  \caption{\small Communication microbenchmark results for comparing only communication performance at different phases of 3D-parallelism, for two different parallelization strategies of \transsmall from~\autoref{fig:Motiv}.}
  \vspace{-1.5em}
  \label{fig:Microbench}
\end{figure}

\autoref{fig:Microbench} presents the communication breakdown across 3D parallelism phases for two parallelization strategies for \transsmall.
For the MP(20)-DP(1)-PP(1) strategy, there are only wafer-wide All-Reduce operations for the MP communication. The baseline effective BW utilization is bounded by the corner NPUs since they have only 2 links to other NPUs. This limits the average network BW utilization of each NPU to be around $2\times 750 \text{GBps} = 1500 \text{GBps}$. In \ours-A, each NPU-L1 BW is 3 TBps, but NPU-L2 BW is 375GBps.\footnote{
Assuming the L1-L2 BW is equally shared among all NPUs.}
Using a similar analysis as \cite{Themis}, we see that hierarchical collectives result in NPU-L2 BW being the bottleneck and the effective NPU BW utilization is $375 \text{GBps} + 4\times 375 \text{GBps} = 1850 \text{GBps}$. In \ours-B, the L1 switches first perform the All-Reduce and then use the entire L1-L2 BW to forward the data to the L2 switches for the second All-Reduce. Therefore, each NPU can send the data to L2 switch at the speed of $1500 \text{GBps}$ (L1-L2 BW). However, since it is an in-network collective execution, the amount of traffic each NPU sends out is almost half of the traffic in the endpoint-based collective. \ours-C has much more L1-L2 BW and therefore each NPU can drive the BW utilization to $3 \text{TBps}$. In \ours-D, an additional in-network collective execution reduces the traffic by half in addition to the $3 \text{TBps}$ NPU BW utilization.


The MP(2)-DP(5)-PP(2) case has all MP (All-Reduce), DP (All-Reduce), and PP (multicast) communications. For the MP communications, the baseline NPU can only utilize 1 link (out of its up to 4 links), resulting in only $750 \text{GBps}$ BW utilization. Since all the communicating NPUs are below the same L1 switch in \ours topologies, they can use the entire $3 \text{TBps}$ of NPU-L1 BW to communicate. Additionally, in the special case when the number of peer NPUs is two, the amount of traffic for endpoint-based vs. in-network execution is the same. Hence, all \ours variants have the same performance for MP communication.

Again, the baseline is limited by the corner NPUs, which can utilize only one of their links for DP communication. Hence, the baseline NPU BW is $750 \text{GBps}$. In \ours, and for the DP communication, each NPU should communicate with four other NPUs under different L1 switches. Therefore, in \ours variants the L1-L2 BW should be shared across four collective flows. Therefore, L1-L2 BW plays a significant role in the performance of this collective. In \ours-A, each NPU has an average NPU-L2 BW of $375 \text{GBps}$, and hence, the NPU BW utilization is only $375 \text{GBps}$, which is worse than the baseline. In \ours-B, however, the L2 switch is used to perform All-Reduce for each flow. This reduces the traffic generated by each NPU roughly by $37.5\%$, which makes its overall performance closer to the baseline. In \ours-C, however, the NPU-L2 BW is increased to $3 \text{TBps}$. Finally, \ours-D Improves the \ours-C by performing in-network collective and reducing the traffic by $37.5\%$.

For the PP comm, the baseline NPU can utilize one of its links to forward data to the next pipeline stage and hence, its BW utilization is 750GBps.
Note that this is possible since in the case of language models such as \transsmall, one NPU within the mp group is sufficient to multicast the output to all NPUs at the next stage,\footnote{
All NPUs within the same MP group produce the same output in this case}
and hence, there is no contention between NPUs of the same MP group at the same stage.  
In \ours, all peer NPUs are below the same L1 switch and can utilize the entire $3 {TBps}$ BW for the PP comm.




\ParaTitle{Discussion: \ours's NPU to L1 Topology Logic.}
Now that we have presented the microbenchmark results, we can discuss why we preferred to choose a tree-based topology to connect every four NPUs to the L1 switches. An alternative solution can be a fully-connected topology to connect every four NPUs and then use only one switch level. However, this design choice still suffers from the endpoint-based effects (i.e., increased use of compute and memory BW at the endpoint) discussed in~\autoref{subsec:simFramework}. Furthermore, as explained in~\autoref{subsec:collectivealgs}, endpoint-based methods produce more communication traffic. For example, in the case of four NPUs, the most endpoint-based BW optimal algorithms produce 1.5D traffic per NPU to perform an All-Reduce of size D \cite{collective1, Themis}, while the in-network collective execution produces only D traffic per NPU \cite{switchml}, 50\% lower than the fully connected topology.


\begin{figure}[h]
  \centering
  \includegraphics[width=0.8\linewidth]{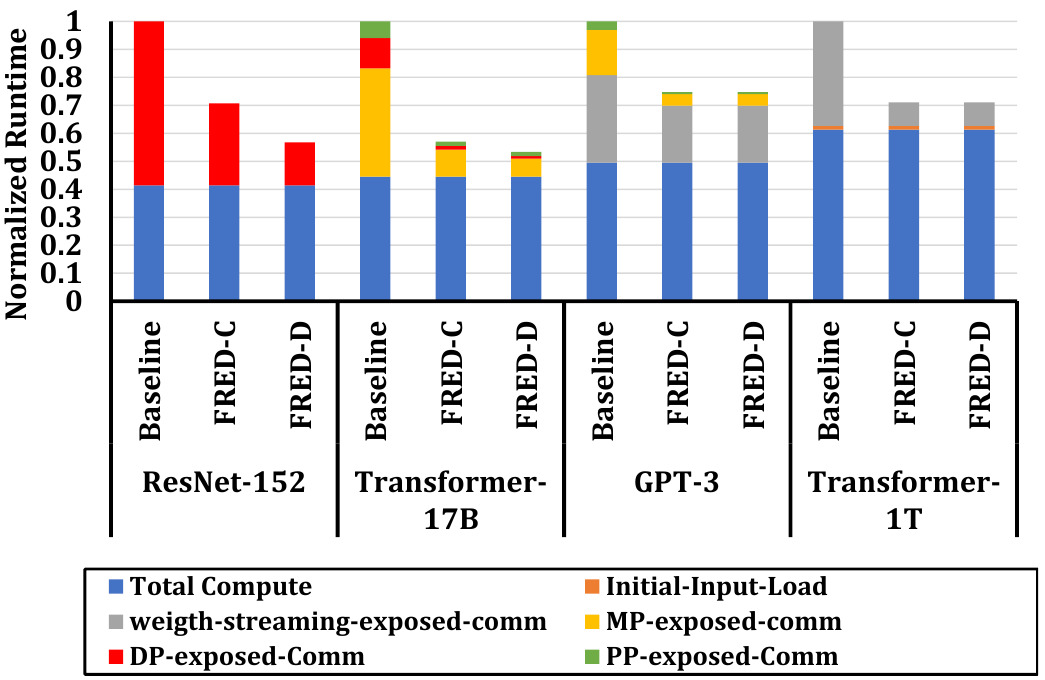}
  \vspace{-1em}
  \caption{\small End-to-end training times are decomposed into compute times and different communication times. The runtime of each workload is normalized to its corresponding baseline.}
  \vspace{-2em}
  \label{fig:results}
\end{figure}

\subsection{Full Workload Results: Deep Dive}\label{subsec:result_deepdive} 
\autoref{fig:results} shows the end-to-end runtimes of the training workloads for the baseline vs. \ours.  
Due to space limitations, we only show the \ours-C and \ours-D in comparison with the baseline. However, we note that \ours-A and \ours-B results are between the baseline and \ours-C, in terms of performance.  
In general, input activations, compared to the model parameters, are relatively small in size and hence, do not have significant overhead on the total iteration time.  
Additionally, the input activations of the next iteration can be prefetched to the wafer whenever the wafer-scale interconnect is idle.  
Hence, we observe \textbf{no} \emph{initial\_input\_load} exposed comm for any of our target workloads, except for the \translarge.

\resnet uses pure DP with a weight stationary model.  
Hence, the only communication costs that repeat on each training iteration are the input minibatch loading and DP communication.  
As explained earlier, in wafer-wide All-Reduce collective, the baseline is able to utilize $1.5$ TBps of NPU BW.  
\ours-C and \ours-D can achieve $3$ TBps NPU BW but \ours-D can further reduce the network traffic by $\approx 2\times$, resulting in a significant reduction of DP exposed comm.  
Thus, \ours-C and \ours-D can improve the end-to-end training runtime by $1.41\times$ and $1.76\times$, respectively, for \resnet.

\transsmall uses all dimensions of the 3D-parallelism and therefore, has all DP, MP, and PP communication overheads.  
The baseline device placement favors MP, but compromises the PP and DP comms, especially due to the non-aligned parallelization strategy dimensions as explained in~\autoref{sec:motivation}.  
Another drawback of the baseline is the underutilized links due to the non-overlapping nature of MP/DP/PP comms (see~\autoref{sec:motivation}).  
\ours-C, on the other hand, does not have the problem of underutilized links and non-aligned parallelization strategies.  
It also does not require favoring any of DP, MP, or PP over the other strategies.  
\ours-D can further improve the MP and DP collectives' performance due to in-switch collective execution capability.  
As a result, \ours-C and \ours-D can improve the overall end-to-end training performance by $1.75\times$ and $1.87\times$, respectively.

\gpt combines weight streaming with 3D-parallelism.  
Using the analysis of~\autoref{sec:motivation}, the baseline topology is unable to stream weights with the full line-rate of I/O controllers.  
The reason is that the hotspot link requires $(2\times 5\:-1)\times 128\text{ GBps}\:=\:1152\text{ GBps}$, while link capacity is only $750$ GBps.  
Therefore, the I/O channels should work with $\frac{750}{1152}=0.65\times$ of the line-rate.  
The MP/PP comm performance of \ours-C and \ours-D is $\approx 4\times$ better than the baseline, due to the underutilized links in the baseline.  
Note that the reason why \ours-C and \ours-D have the same performance for MP collective comm is because dim(MP)=2.  
In this special case, as explained earlier, end-to-end and in-switch collective execution have the same amount of networking traffic and hence, have the same performance.  
In total, \ours-D and \ours-C outperform the baseline by $1.34\times$ in terms of overall training time for \gpt. 

\translarge is another weight streaming workload, but with only DP parallelism.  
As a result, the weight streaming delay is the only communication overhead in addition to the initial input load.  
The high-performance compute NPUs and limited off-chip I/O BW puts the weight streaming performance directly on the critical path.  
This means that the NPUs can work with the line-rate of the weight being streamed, and the main limiting factor is how fast all the weights can be streamed.  
In this case, both \ours-C and \ours-D can leverage the full I/O BW, while the baseline topology can only work with $0.65\times$ of the total I/O BW as explained earlier.  
Additionally, since I/O controllers are always being utilized for weight streaming, there is no idle time to prefetch the input minibatch of the next iteration during the current training iteration.  
Hence, the initial input load cannot be hidden, although its overhead is very negligible.  
In total, using \ours-C/\ours-D improves the training time 
by $1.4\times$.

\begin{figure}[t!]
    \centering
    \begin{subfigure}[b]{0.46\textwidth}
        \centering
        \includegraphics[width=\linewidth]{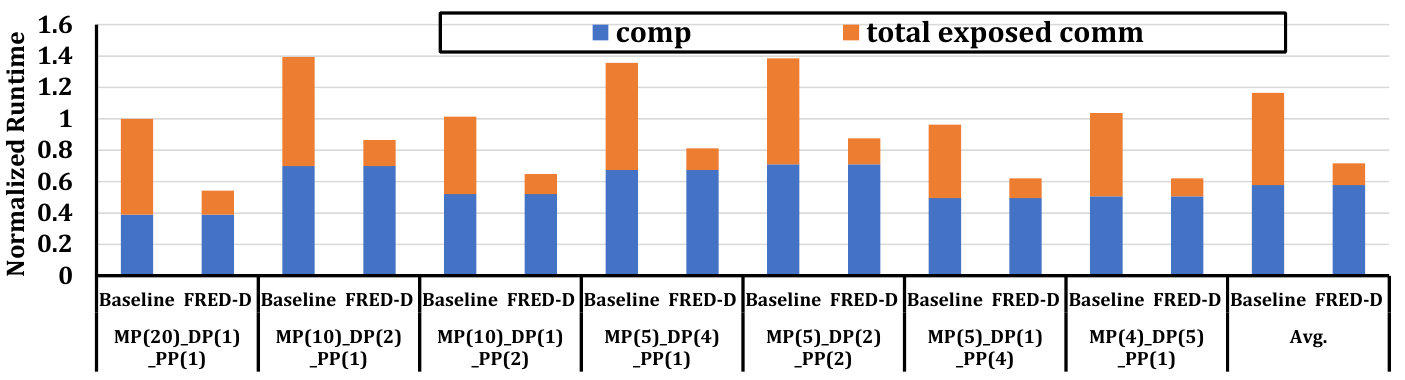} 
        \caption{\transsmall}
        \label{fig:resutls_trans17B_explore}
    \end{subfigure}
    \hfill
    \begin{subfigure}[b]{0.46\textwidth}
        \centering
        \includegraphics[width=\linewidth]{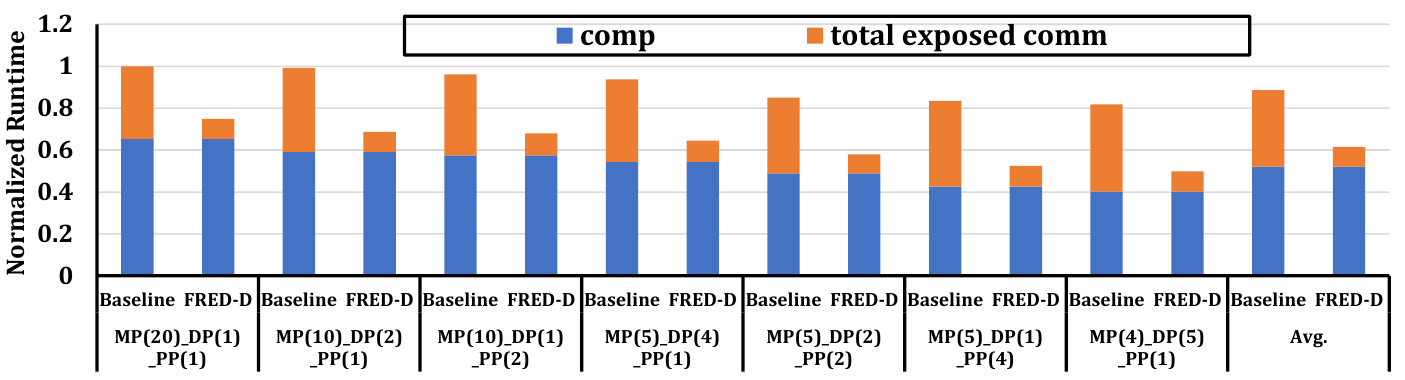} 
        \caption{\translarge}
        \label{fig:resutls_trans1T_explore}
    \end{subfigure}
    \vspace{-1em}
    \caption{Baseline vs. \ours-D for various parallelization strategies of \translarge and \transsmall}
    \label{fig:main_figure}
\end{figure}

\vspace{-1mm}
\subsection{Various Parallelization Strategies}\label{subsec:result_various}
To test the efficiency of \ours for different parallelization strategies, we pick two workloads, \transsmall and \translarge, and compare the baseline performance vs. \ours-D in~\autoref{fig:resutls_trans17B_explore} and~\autoref{fig:resutls_trans1T_explore}, respectively. The Avg. bars are obtained across all parallelization strategies similar to~\autoref{fig:Motiv}, however, not all individual parallelization strategies of~\autoref{fig:Motiv} are shown in~\autoref{fig:resutls_trans17B_explore} and~\autoref{fig:resutls_trans1T_explore} due to lack of space. As can be observed from both figures, \ours-D can significantly improve communication performance and reduce the total exposed communication in all parallelization strategies.

Such improvements make the most compute-efficient (i.e., least compute time) parallelization strategy also to be the be the best parallelization strategy overall. For example, for \transsmall the most compute efficient strategy is MP(20)-DP(1)-PP(1). However, this configuration does not have the lowest overall training time in the baseline system due to its huge exposed communication overheads. Thanks to the benefits of \ours-D in reducing the share of exposed communication overheads, this configuration is now the most optimal compared to other parallelization strategies. This is also true for \translarge, where the most compute-efficient strategy (i.e., MP(5)-DP(1)-PP(4)) is now the most optimal strategy. 

Overall, when averaged across all parallelization strategies, \ours-D can improve the exposed communication time by $4.22\times$ and $3.92\times$, resulting in training speedup by $1.63\times$ and $1.44\times$ for \transsmall and \translarge, respectively.

\ParaTitle{Discussion: going beyond a single wafer.}
While the main focus of this paper is on providing flexible on-wafer interconnects to allow for more flexible parallelization strategies, here we discuss the possible scenarios when the model cannot fit on a single wafer. The first method is to pyramidically load and unload parts of the model (i.e., weight streaming) as we considered and evaluated in the paper. However, in some cases more than one wafer is needed for training to reduce the training time. In that case, the optimal inter-wafer topology is an open question. Some methods use reduction trees to accumulate the gradients obtained from different wafers \cite{Cerebras}. This method, although efficient for data-parallel strategy across the wafers, is not flexible if we consider other parallelization strategies across wafers. A \ours-like inter-wafer interconnect can be constructed to allow for more flexibility across the wafers. In any case, on-wafer \ours\ topology can work in tandem with the inter-wafer interconnect to form efficient hierarchical collectives. For example, a global all-reduce can be broken into: i) a special intra-wafer reduce scatter performed by \ours\ where only the boundary NPUs with access to the I/O maintain the results, followed by ii) an All-Reduce facilitated by the inter-wafer interconnect where boundary NPUs reduce the data across different wafers, iii) followed by the final intra-wafer special All-Gather done by \ours\ where the boundary NPUs broadcast the final result to all NPUs within the same wafer.

\ParaTitle{Discussion: going beyond 3D Parallelism.}
While the main focus of this paper was on the MP/DP/PP parallelism, recently, more parallelization strategies have been proposed. Examples include Expert-Parallelism (EP)~\cite{ExpertParallel}, Context Parallelism (CP)~\cite{ContexParallel}, and more customized and non-homogeneous strategies where the parallelization strategy might change layer by layer~\cite{FlexFlow}. While not quantitatively studied in this paper, we expect that increasing the parallelization strategy dimensions further increases the network congestion and reduces the effective network BW for each parallelism dimension on the baseline 2D Mesh. This highlights the need to have a flexible network fabric such as \ours.  

%% file: content/9_related.tex
\vspace{-2mm}
\section{Related Works}
\label{sec:related}

\ParaTitle{Accelerator Fabrics.}
Prior works on \textit{flexible} DNN Accelerators~\cite{maeri,sigma,flexagon,eyeriss_v2,hojabr2017customizing} have explored indirect topologies such as Benes/Fat-tree/Clos for efficiently distributing operands and reducing partial sums.
This work leverages this concept to build a topology optimized for collectives.

\ParaTitle{In-switch Collectives.}
The idea of in-switch collective execution has been proposed in many previous works for different network levels.
The \emph{P4} language~\cite{P4Language} allows for offloading application-specific tasks to network switches that support the P4 abstract architecture.
\cite{switchml,ATP} proposed programming datacenter Ethernet switches for offloading the All-Reduce collective for data-parallel training.
iSwitch~\cite{CommBottleneck1} utilizes FPGA logic within switches to offload the All-Reduce functionality for the distributed training of reinforcement learning.
Mellanox SHARP~\cite{MellanoxSHARP} is an Infiniband switch architecture for performing collectives.
Klenk \textit{et al.}~\cite{NVidiaSwitch} propose a method to offload collectives to the scale-up (e.g., NVlink\cite{NVlink}) NPU fabric.
Clos topologies have also been explored in prior works~\cite{hojabr2017customizing,ATP}.
A fundamental difference between these works and \ours is that they are proposed for \textit{off-chip} networks, which have significantly less BW compared to on-package networks.
In many of these solutions, the internal switch BW should be at least $2\times$ and $P\times$ the link BW to be efficient (i.e., line-rate) for All-Reduce and Reduce between $P$ ports, respectively.
This is due to the switch architecture that performs the reductions only after the routing and on the output port.
While the difference between off-chip links and on-chip switch architectures allows for provisioning such BW differences, it is not applicable for on-package/on-wafer platforms where the links are on-chip and can have the same BW as the switches.
In contrast, \ours performs the reduction operations in multiple steps ($\mu$Switches) during the routing on the \ours interconnect.
Hence, the \ours switch works with the same BW as the links and can provide line-rate throughput.

%% file: content/10_conclusion.tex
\section{Conclusion \& Future Works}\label{sec:conclusion}

We propose \ours, a high-BW wafer-scale fabric that is flexible for different configurations of the 3D parallelization strategies of distributed training workloads.
\ours is able to support concurrent in-network collective execution efficiently, enabling the upper-level compiler to further optimize the parallelization strategy for compute and memory utilization.
We plan to study \ours for distributed inference as a part of our future work.